\documentclass[journal,11pt,draftcls,onecolumn]{IEEEtran}

\usepackage{cite}

\usepackage{eufrak}
\usepackage[dvips]{graphicx}

\ifCLASSINFOpdf
\else
\fi
\DeclareMathAlphabet{\mathpzc}{OT1}{pzc}{m}{it}

\usepackage[cmex10]{amsmath}
\usepackage[tight,footnotesize]{subfigure}

\usepackage[center,small,font=tt,labelfont=bf]{caption}
\DeclareCaptionFormat{myformat}{\fontfamily{phv}\selectfont#1#2#3\par}  
\makeatletter
\newlength \figwidth
\if@twocolumn
\else
\fi
\makeatother

\begin{document}
\title{Bayesian Estimation of a Gaussian source in Middleton's Class-A Impulsive Noise}
\author{Paolo~Banelli,~\IEEEmembership{Member,~IEEE}}

\thanks{The author is with the Department of Electronic and Information
Engineering, University of Perugia, 06125 Perugia, Italy (e-mail:
paolo.banelli@diei.unipg.it;).}

%
%

\maketitle
\arraycolsep0.2pt
\setcounter{page}{1}

\begin{abstract}
The paper focuses on minimum mean square error (MMSE) Bayesian estimation
for a Gaussian source impaired by additive Middleton's Class-A impulsive noise.
In addition to the optimal Bayesian estimator, the paper considers also
the soft-limiter and the blanker, which are two popular suboptimal estimators
characterized by very low complexity.
The MMSE-optimum thresholds for such suboptimal estimators are
obtained by practical iterative algorithms with fast convergence.
The paper derives also the optimal thresholds according to
a maximum-SNR (MSNR) criterion, and establishes connections with
the MMSE criterion.
Furthermore, closed form analytical expressions are derived for the
MSE and the SNR of all the suboptimal estimators, which perfectly
match simulation results.
Noteworthy, these results can be applied to characterize the
receiving performance of any multicarrier system
impaired by a Gaussian-mixture noise, such as asymmetric
digital subscriber lines (ADSL) and power-line communications (PLC).
\end{abstract}


\section{INTRODUCTION}
\label{sec:introduction}
\textsc{Interference }and noise with impulsive non-Gaussian distributions
may impair the performance of several systems including communications,
controls, sensors and so forth. Middleton has proposed widely accepted
canonical models for interference \cite{Middleton:1, Middleton:1977, Middleton:1983}, which are capable to
characterize ``intelligent'' (e.g., information bearing), as well as
``non-intelligent'' (e.g., natural or man-made) noises.
Although Middleton's noise models were widely investigated to identify the
interference behavior \cite{Middleton:1, Middleton:1977, Middleton:1983, Berry:1981,
Middleton:1999}, to estimate their canonical parameters
\cite{Middleton:1979, Zabin:1989, Zabin:1991, Blackard:1993, Zhong:2007}, and to detect finite
alphabets in digital communications \cite{Rappaport:1966, Spaulding:1977, Spaulding:1985, Middleton:1995, Stein:1995, Maras:2003},
to the best of the author knowledge results for the optimum Bayesian estimator (OBE) of
Gaussian sources in Class-A impulsive noise are still lacking or, at least, not widely acknowledged.
%
Thus, the first aim of the paper is to derive the minimum mean squared error (MMSE) OBE
for a scalar Gaussian source impaired by Middleton's Class-A canonical noise.
Noteworthy, such an OBE is useful also as a preprocessing stage for
estimation and detection algorithms that are designed under AWGN hypotheses and, consequently,
are not robust to impulsive noises \cite{Rappaport:1966, Zhidkov:2006}.
Although the paper derives the analytical expression of the OBE in a closed form,
its use may be restricted in some practical applications due to complexity or
implementation constraints.
For instance, this is the case when the protection from the impulsive
source has to be granted in the analogic domain either to protect the
device, or to limit the input dynamic range of A/D converters.
In these cases it is possible to employ simpler suboptimal devices
that are robust to high noise peaks: a possibility is to resort
to a blanking-nonlinearity (BN) that nulls out the received signal
when it overpasses a given threshold or, alternatively,
to a soft-limiter (SL) that simply clips the signal when it
overpasses the threshold.
In both cases, only the blanking or the clipping
thresholds have to be optimized in a MMSE Bayesian sense.
Actually, although the SL estimator (SLE) and the BN estimator (BNE)
are suboptimal with respect to the OBE, the derivation of the
optimum Bayesian thresholds is analytically much harder than
the computation of the OBE expression.
Anyway, the paper shows that in both cases the computation of the optimum thresholds
can be formulated as the solution of a fixed-point problem \cite{Mayers:2003},
which is proved to always admit a solution, obtainable by standard iterative
approaches with fast convergence.
The comparison of the shape of the OBE curve with the simplified SLE and BNE,
can intuitively illuminate whether the BNE or the SLE is the best
simplified strategy. Typically, the MSE is the
quantitative parameter that is used to choose among different estimators:
the paper shows that the best choice among the SLE and the BNE strictly depends
on the statistical characteristics of the received signal, which are
summarized by the average signal-to-noise power ratio (SNR),
the noise peakness, the average number of emitting noise sources, etc..
Theoretical and simulation results highlight that in almost all the scenarios
(at least) one of the two suboptimal estimators does not suffer any significant
MSE loss with respect to the OBE, further motivating their use.

From a practical application perspective, the OBE and the suboptimal SLE and BNE are
of valuable help in those applications where the quantity of interest can be modeled,
or approximated, by a Gaussian probability density function (\emph{pdf}).
This is the case, due to the central limit theorem (CLT) \cite{Papoulis:1991},
when the quantity of interest is generated by the superposition of several non-dominating
random quantities, as it happens for instance in multicarrier-based communication systems.
In particular, asymmetric digital subscriber lines (ADSL) \cite{Kyees:1995}
and power-line communications (PLC) \cite{Pavlidou:2003}, which are known to face cumbersome
impulsive noise scenarios \cite{Henkel:1995, Zimmermann:2002, Ma:2005, Nassar:2011},
can greatly benefit by employing the proposed estimators at the receiver side.
More generally, the proposed estimators can be used in any multiple-input single-output (MISO)
system with a high number of inputs, which is impaired by impulsive noise.

A different criterion, based on the maximization of the SNR,
has been used in \cite{Zhidkov:2006} and \cite{Zhidkov:2008} to set the optimal
SL and BN thresholds in multicarrier communication systems impaired by impulsive noise.
However, while \cite{Zhidkov:2006} and \cite{Zhidkov:2008} consider a complex Gaussian source,
this paper concentrates on real Gaussian sources (such as those involved in ADSL- and PLC-based communications.
Thus, another contribution of the paper is the derivation of the maximum-SNR (MSNR) thresholds for
the BN and SL of real-valued signals, which are different from those derived in \cite{Zhidkov:2006} and \cite{Zhidkov:2008}.
Note that, while MMSE and MSNR are equivalent in pure AWGN scenarios \cite{Guo:2005}
where the MMSE estimator is linear, this is not the case when the noise
is a Gaussian-mixture, which leads to a non-linear MMSE estimator.
Due to the fact that the MMSE and the MSNR approaches are not equivalent,
they lead to SL and BN suboptimal estimators with different thresholds:
this paper shows when the two thresholds are similar and, conversely, when they are different.
Whether it is better to maximize the SNR or minimize the MSE depends
on the specific application and design constraint. This is not the subject of the paper,
which however establishes also the connection between the MMSE and the MSNR criteria:
by exploiting this connection, the final contribution of the paper
is the derivation of closed form expressions of the MSE and SNR for
the suboptimal SL and BN estimators.

The paper is organized as follows: section \ref{sec:system} introduces the system model,
while the OBE, SLE, and BNE are derived in sections \ref{sec:optimum},
\ref{sec:bayesian}, and \ref{sec:mylabel2}, respectively.
Successively, section \ref{sec:maximum}
concentrates on the MSNR criterion, proposes a method that greatly simplifies its theoretical
computation, and derives the equations to iteratively compute the MSNR-optimal thresholds
for the SLE and the BLE.
Section \ref{sec:SNR-MSE-theory} formally establishes the relationship between MSE and SNR,
and highlights that also the theoretical MSE of the BLE and SLE can be derived
with significant lower computational complexity with respect to a classical approach.
Finally section \ref{sec:computer} is dedicated to computer simulations that confirm
the theoretical findings, while the conclusions are drawn in the last section.

$E\{\cdot\}$ is generally used throughout the paper for statistical expectation, while
$E_X\{\cdot\}$ is used to make explicit that the expectation is computed
with respect to the \emph{pdf} of the random variable $X$. Furthermore,
$g^{(k,\alpha )} (x,n;\alpha )$ is used for the $k\textrm{-th}$
derivative of $g(x,n;\alpha )$ with respect to $\alpha $.

\section{System Model}
\label{sec:system}
\begin{figure}[ht]
\centerline{\includegraphics[trim=0cm 1cm 0cm 0cm, clip=true, width=\figwidth]{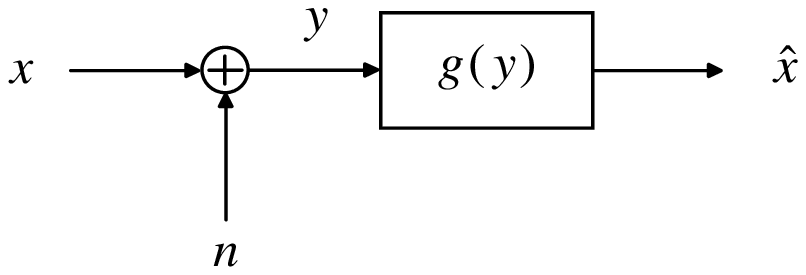}}
\caption{System model}
\label{SystemModel}
\end{figure}
Let's consider a zero-mean real source $x$ with average power $\sigma_X^2 $ and Gaussian \emph{pdf}
$f_X (x)=G(x;\sigma_X^2 )= (\sqrt{2\pi}\sigma_{X})^{-1}e^{-{x^2}/{2\sigma_X^2 }}$,
impaired by a Class-A impulsive noise $n$ with average power
$\sigma_N^2 $, as shown in \figurename~\ref{SystemModel} and summarized by
\begin{equation}
\label{eq:sig-plus-noise}
y=x+n.
\end{equation}
The Class-A impulsive noise subsumes also the presence of a background
zero-mean thermal AWGN $n_t$, with average power $\sigma_t^2$.
Specifically, the impulsive noise \emph{pdf} is a Gaussian-mixture expressed by
\begin{equation}
\label{eq:noise_pdf}
f_N (n)=
\sum\limits_{m=0}^\infty {\beta _m G(n;\sigma _m^2 )}
=\sum\limits_{m=0}^\infty \frac{\beta _m }{\sqrt {2\pi \sigma
_m^2 } }e^{-\frac{n^2}{2\sigma_m^2 }},
\end{equation}
where the weights $\beta_m=e^{-A}A^m/m!$ represent the Poisson-distributed
probability that $m$ noise sources simultaneously contribute to the impulsive event
\cite{Middleton:1977, Berry:1981}. The power $\sigma_m^2$
associated to the simultaneous emission from $m$ noise sources is
expressed by
\begin{equation}
\label{eq:sigma2_N}
\sigma _m^2 =\frac{m/A+T}{1+T}\sigma_N^2 =m\frac{\sigma _I^2 }{A}+\sigma
_t^2 ,
\end{equation}
where $\sigma_N^2 =E\{n^2\}=\sigma_I^2 +\sigma_t^2 $, $\sigma_I^2$
represents the impulsive part of the noise power, $T=\sigma _t^2 /\sigma
_I^2$ is the power ratio among the AWGN and the impulsive part of the
noise $n$, and $A=E\{m\}=\sum\limits_{m=0}^\infty {m\beta _m }$ represents
the average number of impulsive sources that are simultaneously active.
The three canonical parameters $A$, $T$, and $\sigma_N^2 $ completely  specify
the statistical structure of the  Class-A noise. In particular,
low values of $A$ identify rare and highly-peaked impulsive sources,
while conversely high values of $A$ makes the impulsive noise more
similar to an AWGN, by a CLT argument. The interested readers are redirected
to \cite{Middleton:1977, Middleton:1999, Berry:1981} and references therein
for further details on the Class-A model, and to
\cite{Middleton:1979, Zabin:1989, Zabin:1991, Blackard:1993, Zhong:2007}
for the estimation of the canonical parameters $A$, $T$, and $\sigma_N^2 $.
This paper assumes that the
canonical parameters have been perfectly estimated by the observing system.

\section{Optimum Bayesian Estimator (OBE)}
\label{sec:optimum}
\begin{figure}[t]
\centering %
\subfigure[{$T=1$, $\sigma_X^2 =10$, $\sigma_N^2 =1$}
\label{fig2}]{\includegraphics[trim=0.2cm 0cm 0.5cm 0cm, clip=true, width=0.9\figwidth]{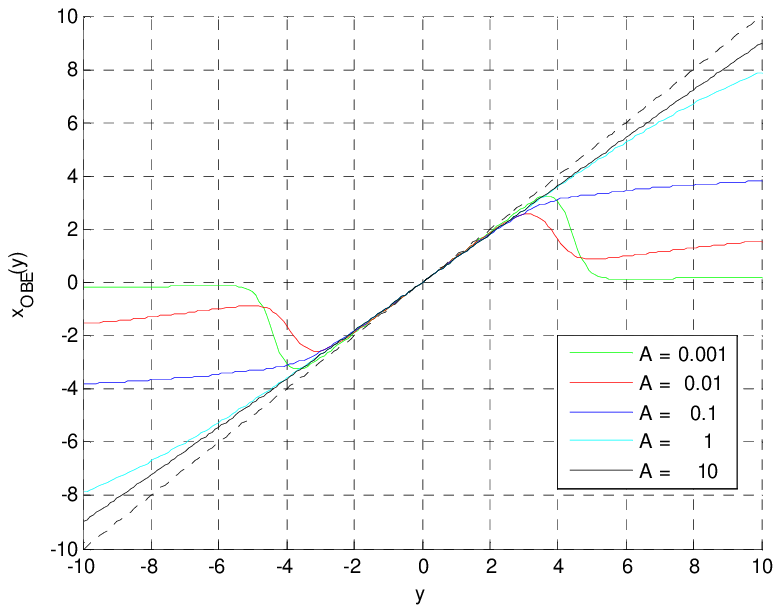}}
\subfigure[{$T=1$, $\sigma_X^2 =\sigma_N^2 =1$}
\label{fig3}]{\includegraphics[trim=0.2cm 0cm 0.5cm 0cm, clip=true, width=0.9\figwidth]{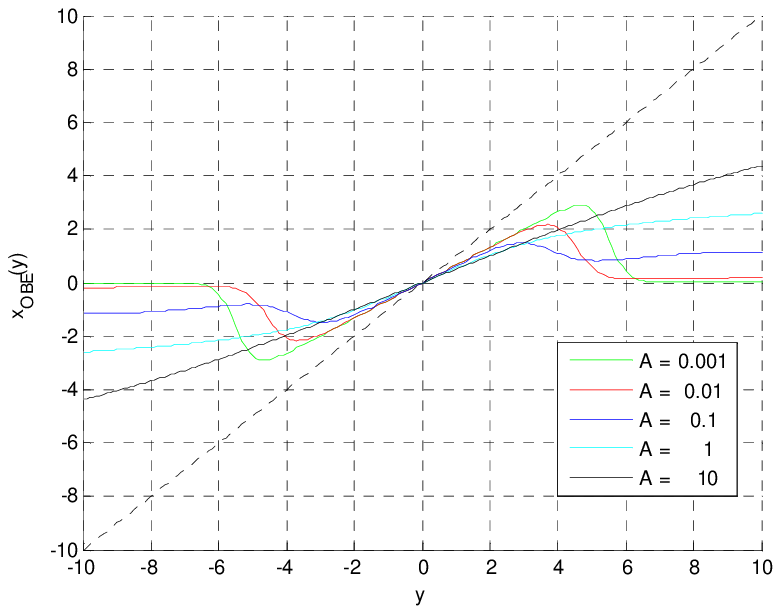}}%
\caption{OBE for several values of $A$ \label{fig2fig3}}
\end{figure}

The optimum MMSE Bayesian estimator of $x$ given the observed
data $y$, is expressed by \cite{Kay:1993}
\begin{equation}
\label{eq:OBE_def}
\hat {x}_{\textrm{OBE}} (y)=E_{X\vert Y} \{x\vert y\}=\int\limits_{-\infty }^{+\infty
} {xf_{X\vert Y} (x;y)dx} ,
\end{equation}
where $f_{X\vert Y} (x;y)$ represents the posterior \emph{pdf} of the source
$x$ for a given observation $y$. By exploiting Bayes rules for conditional
\emph{pdf}s, (\ref{eq:OBE_def}) can be expressed as
\begin{equation}
\label{eq:OBE_def2}
\hat {x}_{\textrm{OBE}} (y)=\frac{1}{f_Y (y)}\int\limits_{-\infty }^\infty x
f_{Y\left| X \right.} \left( {y;x} \right)f_X (x)dx.
\end{equation}
Due to the fact that the impulsive noise $n$ is independent of $x$,
it is well known \cite{Papoulis:1991} that the \emph{pdf} of the
observed value $y$ in (\ref{eq:sig-plus-noise}) is given by the
linear convolution of the
Gaussian \emph{pdf} $f_X (y)$
with the Middleton-A \emph{pdf} $f_N(y)$ in (\ref{eq:noise_pdf}),
as expressed by
%
\begin{equation}
\label{eq:pdf_convolution}
\begin{array}{rcl}
f_Y \left( y \right)
&=&\sum\limits_{m=0}^\infty {\beta _m G(y;\sigma _m^2 )}\ast G(y;\sigma_X^2 )
\\
&=&\sum\limits_{m=0}^\infty {\frac{\beta _m }{\sqrt {2\pi
(\sigma _m^2 +\sigma_X^2 )} }e^{-\frac{y^2}{2\left( {\sigma _m^2 +\sigma_X^2 } \right)}}},
\end{array}
\end{equation}
where $\ast$ stands for the linear convolution operator and it is
exploited that the convolution of two Gaussian \emph{pdf}s
generates a new Gaussian \emph{pdf} \cite{Papoulis:1991}
with a variance equal to the sum of the two original variances.
Observing (\ref{eq:sig-plus-noise}) it is also evident that
$f_{Y\vert X} (y;x)$ in (\ref{eq:OBE_def2}) is expressed by
$f_{Y\left| X \right.} (y;x)=f_N (y-x)$, which plugged in (\ref{eq:OBE_def2}) leads to
\begin{equation}
\label{eq:OBE_conv}
\begin{array}{c}
 \hat {x}_{\textrm{OBE}} (y)=\frac{1}{f_Y (y)}\sum\limits_{m=0}^\infty {\beta _m
\int\limits_{-\infty }^\infty x f_X (x)
G\left(y-x;\sigma_m^2\right)dx}
 \end{array}
\end{equation}
As detailed in Appendix \ref{App:OBE}, the convolution in (\ref{eq:OBE_conv}) between the $m\textrm{-th}$
Gaussian \emph{pdf} $f_m (x)=G(x;\sigma_m^2)$ and $p(x)=xf_X (x)$
can be computed in the Fourier domain by exploiting the properties
of the Fourier transform (FT), leading to
\begin{equation}
\label{eq:OBE_final}
\hat {x}_{\textrm{OBE}} (y)= \sigma_X^2 \frac{\sum\limits_{m=0}^\infty {\frac{\beta
_m }{\left( {\sigma _m^2 +\sigma_X^2 } \right)^{3/2}}}
e^{-\frac{y^2}{2(\sigma _m^2 +\sigma_X^2 )}}}{\sum\limits_{m=0}^\infty
{\frac{\beta _m }{\left( {\sigma _m^2 +\sigma_X^2 }
\right)^{1/2}}e^{-\frac{y^2}{2(\sigma _m^2 +\sigma_X^2 )}}} }\textrm{ }y.
\end{equation}
Equation (\ref{eq:OBE_final}) highlights how the OBE depends on the source average power
$\sigma_X^2$ and the noise canonical parameters $A$, $T$, and $\sigma_N^2$,
through $\beta _m $ and $\sigma _m^2 $ in (\ref{eq:sigma2_N}). The input-output
characteristic of the OBE is plotted in \figurename~\ref{fig2} and
\figurename~\ref{fig3} for several values of the parameter $A$ that
controls the peakness of the impulsive noise \cite{Berry:1981}; it is evident that for very high values of $A$,
when $f_N (n)$ tends to a zero-mean Gaussian \emph{pdf}, the OBE tends to the well
known linear-MMSE estimator, expressed by \cite{Kay:1993}
\begin{equation}
\label{eq:OBE_lin}
\hat {x}_{\textrm{OBE}}^{(lin)} (y)=\frac{E\{xy\}}{E\{y^2\}}y=\frac{\sigma_X^2
}{\sigma_X^2 +\sigma_N^2 }y.
\end{equation}
For lower values of $A$, when the noise is characterizes by rare
and highly peaked impulses, the optimum estimator shows a highly non-linear
nature, by roughly limiting $(A=0.1)$ or blanking $(A=0.001)$ the observed
values $y$ that overpass certain thresholds. The OBE expression in
(\ref{eq:OBE_final}) immediately applies also to any other Gaussian-mixture noise $n$
with a finite number $M$ of mixtures, such that $\beta_m \ge 0$, and $\sum\limits_{m=0}^{M-1} {\beta _m
=1}$.

%

\section{Bayesian Soft Limiter Estimator (SLE)}
\label{sec:bayesian}
\begin{figure}[ht]
\centerline{\includegraphics[trim=0cm 1cm 0cm 0cm, clip=true, width=\figwidth]{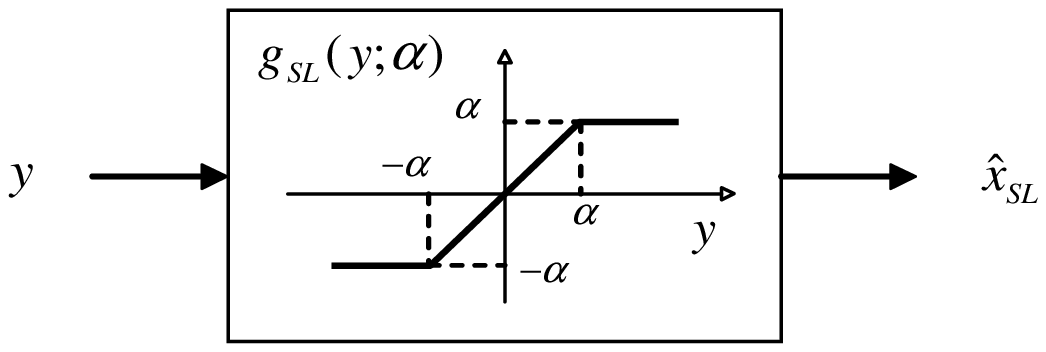}}
\caption{Soft limiter estimator (SLE)}
\label{fig:SoftLimiter}
\end{figure}

The OBE rather involved analytical expression (\ref{eq:OBE_final}) could prevent its use in real-time applications
due to either memory or computational complexity constraints.
This is especially true when the OBE analytical expression is
requested to be adaptive with respect to changes of the source
and of the noise statistical parameters (e.g., the average powers
$\sigma_X^2$ and $\sigma_N^2 $, and the noise peakness factors $A$, and $T)$.
Moreover, it could be also necessary to contrast the impulsive noise
in the analog domain (e.g., before A/D conversion) making OBE
implementations even more challenging.
For these reasons, this section investigates a simpler suboptimum
estimator, namely the SLE shown in \figurename~\ref{fig:SoftLimiter},
which
is typically employed to contrast impulsive noise \cite{Zhidkov:2006} adding
robustness to the system by clipping the signal values that
exceed a given threshold $\alpha $.
Thus, the only parameter to optimize in the Bayesian sense is the clipping threshold $\alpha $,
which obviously would depend on the noise parameters $A$, $T$, and $\sigma_N^2$, as well as on
the source power $\sigma_X^2 $. Meaningfulness of such an optimization,
which leads to the SLE, is also suggested by the OBE shapes in
\figurename~\ref{fig2} and \figurename~\ref{fig3}, which
for certain noise parameters (e.g., $A=0.1)$ resembles the SLE of \figurename~\ref{fig:SoftLimiter}.
The output of the SLE is expressed by a non-linear input-output
characteristic $\hat {x}_{\textrm{SL}} =g_{\textrm{SL}} (y;\alpha )$:
thus, the SLE estimation error $e_{\textrm{SL}}$ depends on the selected
threshold $\alpha$, as well as on the statistical
properties of the source $x$ and the noise $n$. This is expressed by
\begin{equation}
\label{eq:e_SL}
e_{\textrm{SL}}=x-g_{\textrm{SL}} (x+n;\alpha )=h_{\textrm{SL}} (x,n;\alpha )=\left\{\begin{array}{ll}
 x+\alpha 	\quad &, x+n<-\alpha \\
 -n	\quad &, \vert x+n\vert \le \alpha \\
 x-\alpha 	\quad &, x+n>\alpha
\end{array}.\right.
\end{equation}
The SLE estimator is defined by selecting $\alpha _{\textrm{SL}}^{(mse)} $ according
to the MMSE criterion, as expressed by
\begin{equation}
\label{eq:alphaSL_min}
\alpha _{\textrm{SL}}^{(mse)} =\mathop {\arg \min }\limits_{\alpha  \in {\cal R}^+} \left[
J_{\textrm{SL}}(\alpha) \right]=\mathop {\arg \min }\limits_{\alpha  \in {\cal R}^+}
\left[ {E\left\{ {h_{\textrm{SL}}^2 (x,n;\alpha )} \right\}} \right],
\end{equation}
where $J_{\textrm{SL}}(\alpha)=E\{e_{\textrm{SL}}^2\}$ is the MSE cost function.
Thus, in order to find $\alpha _{\textrm{SL}}^{(mse)} $ it is necessary to solve
\begin{equation}
\label{eq:der_alfaSL}
\begin{array}{rcl}
 J_{\textrm{SL}}^{(1,\alpha)}(\alpha)&=&\frac{\partial }{\partial \alpha }E\left\{ {h_{\textrm{SL}}^2 (x,n;\alpha )}
\right\}=E\left\{ {\frac{\partial }{\partial \alpha }\left[ {h_{\textrm{SL}}^2
(x,n;\alpha )} \right]} \right\} \\
 &=&E\left\{ {2h_{\textrm{SL}} (x,n;\alpha )h_{\textrm{SL}}^{(1,\alpha )} (x,n;\alpha )}
\right\}=0,
 \end{array}
\end{equation}
leading to the following integral equation
\begin{equation}
\label{eq:SL_integral}
\int\limits_{-\infty }^{+\infty } {\int\limits_{-\infty }^{+\infty } {h_{\textrm{SL}}
(x,n;\alpha )h_{\textrm{SL}}^{(1,\alpha )} (x,n;\alpha )f_X (x)f_N (n)dxdn=0} }.
\end{equation}
By substituting (\ref{eq:noise_pdf}), (\ref{eq:e_SL}) and its partial derivative in (\ref{eq:SL_integral}),
Appendix \ref{App:SLE} proves that $\alpha_{\textrm{SL}}^{(mse)}$
is the solution of the following fixed-point equation
\begin{equation}
\label{eq:alfaSL_FixPoint_mse}
\alpha =F_{\textrm{SL}}^{(mse)} (\alpha )=2\sigma_X^2 \frac{\sum\limits_{m=0}^\infty
{\frac{\beta _m }{\sqrt {2\pi (\sigma_X^2 +\sigma _m^2 )} }e^{-\alpha
^2/2(\sigma_X^2 +\sigma _m^2 )}} }{1-\sum\limits_{m=0}^\infty {\beta _m
\textrm{erf}\left( {\alpha /\sqrt {2(\sigma_X^2 +\sigma _m^2 )} } \right)} },
\end{equation}
which always admits a solution.
Moreover, Appendix \ref{App:SLE} also proves that $J_{\textrm{SL}}(\alpha)$
in (\ref{eq:alphaSL_min}) is locally convex for $\alpha \in [0,2.05\sigma_X]$,
which is equivalent to prove that locally $F_{\textrm{SL}}^{(mse)} (\alpha )$
is a contraction mapping \cite{Mayers:2003}.
For this reason, any numerical solution of (\ref{eq:der_alfaSL}) that starts from
$\alpha_0 \in [0, 2.05\sigma_X]$ would converge to the MSE minimum,
as well as the succession $\alpha _{n+1} =F_{\textrm{SL}}^{(mse)} (\alpha_n )$
converges to the exact fixed point solution
$\alpha _{\textrm{SL}}^{(mse)} $ when $n$ goes to infinity \cite{Mayers:2003}.
Thus, $\alpha _{\textrm{SL}}^{(mse)} $can be
numerically approximated by the following iterative algorithm
\begin{tabbing}
{\it A1:} \= {\it Iterative algorithm for optimal SL threshold}\\
1.\> $\textrm{set }\alpha _0 =F_{\textrm{SL}}^{(mse)} (0)\textrm{ and }n=0\textrm{;}$ \\
2.\> $\textrm{whi}$\=$\textrm{le}$  $\vert F_{\textrm{SL}}^{(mse)} (\alpha _n )-\alpha _n \vert>\varepsilon \textrm{ and }n\le n_{\textrm{max}}$ \\
3.\> \> $\alpha _{n+1} =F_{\textrm{SL}}^{(mse)} (\alpha _n )\textrm{ ;}$ \\
4. \> \> $n=n+1;$ \\
5. \> $\textrm{end}$ \\
6. \> $\textrm{set }\alpha _{\textrm{SL}}^{(mse)} =\alpha _n \textrm{.}$
\end{tabbing}
In algorithm A1, $\varepsilon $ represents the accuracy that is requested to
the approximated solution to stop within $n_{\textrm{max}}$ iterations.
Obviously, other iterative numerical approaches can also be used
to solve (\ref{eq:alfaSL_FixPoint_mse}), such as the Newton--Rapson method \cite{Mayers:2003}
to find the root of $F_{\textrm{SL}}^{(mse)} (\alpha )-\alpha =0$.
Note that the local convexity of $J_{\textrm{SL}}(\alpha)$ is also confirmed by
the MSE plots in Section \ref{sec:computer}: thus,
it makes sense that for increasing $\alpha$ (starting from $0$) the
first minimum reached in (\ref{eq:alphaSL_min}) by the
iterative algorithm is also the optimal solution,
as confirmed by all the simulation results in Section \ref{sec:computer}.

\section{Bayesian Blanking Nonlinearity Estimator (BNE)}
\label{sec:mylabel2}
\begin{figure}[ht]
\centerline{\includegraphics[trim=0cm 2cm 1.5cm 0cm, clip=true, width=\figwidth]{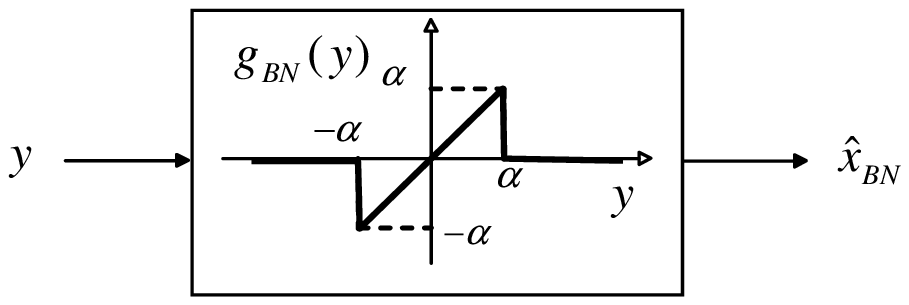}}
\caption{. Blanking nonlinearity estimator}
\label{Fig:Blanker}
\end{figure}

\figurename~\ref{fig2} and \figurename~\ref{fig3} suggest
that for highly impulsive noise behaviors (e.g., $A=0.001)$, the
OBE shape resembles the BN shown in \figurename~\ref{Fig:Blanker},
where the received signal is simply blanked to zero when
its absolute magnitude overpasses the threshold $\alpha $.
Similarly to the SLE, the goal is to derive the optimum MMSE
threshold $\alpha _{\textrm{BN}}^{(mse)} $ for the BN.
In this case the estimation error is expressed by
\begin{equation}
\label{eq:e_BN}
e_{\textrm{BN}} =x-g_{\textrm{BN}} (x+n;\alpha )=h_{\textrm{BN}} (x,n;\alpha )=\left\{
{\begin{array}{lr}
 -n	\quad &,\vert x+n\vert \le \alpha \\
 x	\quad &,\vert x+n\vert >\alpha
 \end{array}} \right.
\end{equation}
and the MSE $J_{\textrm{BN}}(\alpha)$ is expressed by
\begin{equation}
\label{eq:J_BN}
J_{\textrm{BN}}(\alpha) = E\{h_{\textrm{BN}}^2(x,n;\alpha)\}.
\end{equation}
Thus, as detailed in Appendix \ref{App:BNE}, the optimum
$\alpha _{\textrm{BN}}^{(mse)}$ is given by the
solution of $J_{\textrm{BN}}^{(1,\alpha)}(\alpha) =0$, which
is equivalent to the solution of the following fixed-point equation
\begin{equation}
\label{eq:alfaBN_FixPoint_mse}
\alpha =F_{\textrm{BN}}^{(mse)} (\alpha )=\frac{\sum\limits_{m=0}^\infty
{\frac{A^m}{m!}\frac{2\sigma _m^2 }{(\sigma_X^2 +\sigma _m^2 )^{3/2}}\alpha
e^{-\alpha ^2/2(\sigma_X^2 +\sigma _m^2 )}} }{\sum\limits_{m=0}^\infty
{\frac{A^m}{m!}\frac{1}{(\sigma_X^2 +\sigma _m^2 )^{1/2}}e^{-\alpha
^2/2(\sigma_X^2 +\sigma _m^2 )}}}.
\end{equation}
Although the fixed point problem admits a unique (non trivial) solution, as detailed in
Appendix \ref{App:BNE}, $F_{\textrm{BN}}^{(mse)} (\alpha )$ is a monotonically
increasing function and it is not a contraction mapping \cite{Mayers:2003}.
Consequently, $F_{\textrm{BN}}^{(mse)} (\alpha )$ in (\ref{eq:alfaBN_FixPoint_mse})
is not an attraction for the iterative algorithm A1, and an iterative algorithm
that converges to the fixed point is
\begin{tabbing}
{\it A2:} \= {\it Iterative algorithm for optimal} BN {\it threshold}\\
1. \> $\textrm{set } \alpha_0 > 0,\textrm{ } 0<\mu<1, \textrm{ } n=0 \textrm{;}$ \\
2. \> $\textrm{whi}$\=$\textrm{le }$ \= $\vert F_{\textrm{BN}}^{(mse)} (\alpha_n )-\alpha_n \vert>\varepsilon \textrm{ and }n\le n_{\textrm{max}}$ \\
3. \> \> $\alpha_{n+1} =\alpha_n +\mu (\alpha_n -F_{\textrm{BN}}^{(mse)} (\alpha_n ));$ \\
4. \> \> $n=n+1;$ \\
5. \> $\textrm{end}$ \\
6. \> $\textrm{set } \alpha_{\textrm{BN}}^{(mse)} =\alpha_n \textrm{.}$
\end{tabbing}
Differently from A1, in algorithm A2 the starting point $\alpha _0$
has to be greater than zero (for instance $\alpha _0 =\sigma_X)$
to avoid the trivial solution $\alpha =0$, while $\mu $ controls the speed of convergence.

\section{Maximum SNR (MSNR) Estimators}
\label{sec:maximum}
This section is dedicated to introduce the MSNR as an alternative criterion
to optimally design the estimators. This criterion is typically employed in
communication systems, such as ADSL and PLC, where the bit error rate (BER)
performance depends on the SNR rather than on the MSE (\cite{Proakis:1987}).
In this view, \cite{Zhidkov:2006} and \cite{Zhidkov:2008} introduce MSNR
estimators for complex Gaussian process (OFDM signals)corrupted by
impulsive Gaussian mixture noise.
First of all, lets observe that any estimator $\hat {x}(y)$ is in general
obtained as a non-linear transformation $g(y)$ of the observation $y=x+n$.
Anyway, the estimator output can be always decomposed as a
scaled version of the input plus a distortion term $w_y $, as expressed by
\begin{equation}
\label{eq:x_hat_kY}
\hat {x}=g\left( y \right)=k_y y+w_y ,
\end{equation}
where $k_y =E_Y \{g\left(y\right)y\}/E_Y \{y^2\}$
is the linear regression coefficient that grants the distortion term $w_y $
is orthogonal to the input term, i.e., $E\{yw_y \}=0$.
However, due to the presence of the impulsive noise $n$, the
non-linearity input $y=x+n$ does not contain only the
useful information $x$.
Thus, in order to define a meaningful SNR \cite{Zhidkov:2006},
it is more convenient to express the estimator output as
\begin{equation}
\label{eq:x_hat_kX}
\hat {x}(y)=g(y)=k_x x+w_x ,
\end{equation}
where
\begin{equation}
\label{eq:kX_def}
k_x =E_{YX} \{g(y)x\}/E_X \{x^2\}=E_{XN} \{g(x+n)x\}/E_X \{x^2\},
\end{equation}
is the $partial$ linear regression coefficient that grants the distortion
noise $w_x $ is orthogonal to $x$, as summarized by $E_{XW} \{xw_x \}=0$.
Although in general $k_x \ne k_y $, when the inputs are both zero-mean Gaussian
and independent it holds true that $k_x =k_y$
\cite{Banelli:2011IT}.
Proceeding as suggested in \cite{Zhidkov:2006}, the SNR is expressed by
\begin{equation}
\label{eq:SNRout_def}
\begin{array}{rcl}
SNR
&=&\frac{k_x^2 E_X \{x^2\}}{E_{W_x}\{w_x^2\}}=\frac{k_x^2 E_X \{x^2\}}{E_{\hat
{X}} \{\hat {x}^2\}-k_x^2 E_X \{x^2\}}
\\
&=&\frac{1}{E_Y \left\{g(y)^2\right\}/k_x^2 \sigma_X^2 -1},
\end{array}
\end{equation}
where the second equality in (\ref{eq:SNRout_def}) is granted by the uncorrelation between
the useful part and the distortion noise. Thus, in the MSNR sense,
the optimum non-linear estimator is defined by
\begin{equation}
\label{eq:x_hat_SNR_def}
\hat {x}_{SNR} (y)=\mathop {\arg \max }\limits_{g(y)} \left[ {SNR}
\right]=\mathop {\arg \min }\limits_{g(y)}
\left[ \frac{E_Y \{g(y)^2\}}{E_{XY}^2\{g(y)x\}}
 \right].
\end{equation}
In the problem at hand, taking into account that $n$ is distributed
according to (\ref{eq:noise_pdf}), the computation of $k_x $
in the denominator of (\ref{eq:SNRout_def}) and
(\ref{eq:x_hat_SNR_def}) can be obtained by
\begin{equation}
\label{eq:kX_sum}
k_x =\frac{E_{XN} \{g(x+n)x\}}{\sigma_X^2 }=\sum\limits_{m=0}^\infty {\beta
_m \frac{E_{XN_m} \{g(x+n)x\}}{\sigma_X^2 }} ,
\end{equation}
where the subscript $N_m$ means that the expected value is computed
with respect to the $m\textrm{-th}$ Gaussian \emph{pdf} $G(n;\sigma _m^2 )$
associated to the Class-A Gaussian mixture.
This fact suggests that the constant $k_x$ can be expressed
as the weighted sum of other constants
\begin{equation}
\label{eq:kXm_def}
k_{x,m} =\frac{E_{XN_m } \{g(x+n_m )x\}}{\sigma_X^2 },
\end{equation}
which can be interpreted as the gain associated to the ({\it virtual}) useful
components at the output of the non-linear estimator when it is separately
excited by the ({\it virtual}) input $y_m =x+n_m $.
Each {\it virtual} input $y_m $ is the sum of two zero-mean
independent Gaussian random variables and it is also zero-mean
Gaussian distributed with variance $\sigma _{y,m}^2 =\sigma_X^2 +\sigma
_m^2 $.
In this virtual set-up, (see Theorem 1 in \cite{Banelli:2011IT}), it
holds true that
\begin{equation}
\label{eq:kXm_kYm}
k_{x,m} =\frac{E_{XN_m } \{g(x+n_m )x\}}{\sigma_X^2 }=k_{y,m} =\frac{E_{Y_m
} \{g(y_m )y_m \}}{\sigma _{y,m}^2 },
\end{equation}
where the expectation on the right-hand side of (\ref{eq:kXm_kYm}) involves a
single-folded integral, which is much simpler to compute than the two-folded
integral in its left-hand side. Moreover, last equality in (\ref{eq:kXm_kYm}) is
attractive because it lets to exploit widely known results for the
output of several non-linear devices (such as the BN and the SL) excited by
Gaussian inputs
\cite{Rowe:1982, Bussgang:1952, Baum:1957, Davenport:1958, Banelli:2000}.
Similarly, it is straightforward to derive that the average estimator output
power in the numerator of (\ref{eq:x_hat_SNR_def}) can be expressed as
\begin{equation}
\label{eq:Px_hat}
P_{\hat {X}} =E_Y \{g(y)^2\}=\sum\limits_{m=0}^\infty {\beta _m E_{Y_m }
\{g^2(y_m )\}} ,
\end{equation}
where also right-hand side of (\ref{eq:Px_hat}) can exploit results widely available in
the said technical literature for non-linear distortions
of Gaussian random variables.
However, despite the above simplifications, the solution of the functional
optimization problem in (\ref{eq:x_hat_SNR_def}) is not easy and
the derivation of the optimum estimator in the MSNR sense is still an open problem.

Conversely, it is possible to exploit (\ref{eq:kXm_kYm}) and (\ref{eq:Px_hat})
if $g(y)$ is constrained to belong to families of suboptimal estimators
\begin{equation}
\label{eq32}
\hat {x}_{\textrm{XX}} (y)=g_{\textrm{XX}} (y;\alpha ),
\end{equation}
where XX stands for either the SL or the BN, and $\alpha $ is a scalar
parameter that univocally specifies $g_{\textrm{XX}} (\cdot;\alpha )$.
In this case the problem reduces to a classical optimization
with respect to the scalar parameter $\alpha $,
where the optimum MSNR thresholds are expressed by
\begin{equation}
\label{eq:alfa_XX_msnr_def}
\alpha _{\textrm{XX}}^{(snr)} =\mathop {\arg \min }\limits_{\alpha  \in {\cal R}^+} \left[ {E_Y \{\hat
{x}_{\textrm{XX}}^2 (y)\}/(k_x^{\textrm{(XX)}} )^2} \right].
\end{equation}
Thus, taking into account that the logarithm does not change the position of extreme values,
the MSNR threshold is obtained by solving
$ \frac{\partial }{\partial \alpha }\left\{ {\log \frac{E_Y \{\hat {x}_{\textrm{XX}}^2
(y)\}}{(k_x^{\textrm{(XX)}} )^2}} \right\}=0$,
which leads to
\begin{equation}
\label{eq:x_SLBN_def_der}
\frac{1}{E_Y \{\hat {x}_{\textrm{XX}}^2 (y)\}}\frac{\partial }{\partial \alpha }E_Y
\{\hat {x}_{\textrm{XX}}^2 (y)\}-\frac{2}{k_x^{\textrm{(XX)}} }\frac{\partial }{\partial
\alpha }k_x^{\textrm{(XX)}} =0.
\end{equation}
When the non linear device is the SL $g_{\textrm{SL}} (y;\alpha )$ of \figurename~\ref{fig:SoftLimiter},
it is straightforward to derive that
$k_{y,m}^{\textrm{(SL)}}=\textrm{erf} \left(\frac{\alpha }{\sqrt 2 \sigma _{y,m}}\right)$ \cite{Rowe:1982}
and consequently
\begin{equation}
\label{eq37}
k_x^{\textrm{(SL)}} =\sum\limits_{m=0}^\infty {\beta _m \textrm{erf}\left( {\frac{\alpha
}{\sqrt {2(\sigma_X^2 +\sigma _m^2 )} }} \right)},
\end{equation}
where the error function is defined as $\textrm{erf}(x)=(2/\sqrt{\pi})\int_{0}^{x}{e^{-t^2}dt}$.
Analogously, when the non-linear estimator is the BN $g_{\textrm{BN}} (y;\alpha )$ of
\figurename~\ref{Fig:Blanker}, by standard integration techniques it is possible to prove that
\begin{equation}
\label{eq38}
k_m^{\textrm{(BN)}} =\frac{E_{Y_m } \{g_{\textrm{BN}} (y_m ;\alpha )y_m \}}{\sigma _{y,m}^2
}=k_m^{\textrm{(SL)}} -\frac{2}{\sqrt \pi }\frac{\alpha }{\sigma _{y,m}
}e^{-\frac{\alpha ^2}{2\sigma _{y,m}^2 }},
\end{equation}
and, consequently,
\begin{equation}
\label{eq39}
k_x^{\textrm{(BN)}} =k_x^{\textrm{(SL)}} -\frac{2}{\sqrt \pi }\sum\limits_{m=0}^\infty {\beta
_m \frac{\alpha }{\sigma _{y,m}}e^{-\frac{\alpha ^2}{2(\sigma_X^2 +\sigma
_m^2 )}}}. 
\end{equation}
Similarly, plugging $g_{\textrm{SL}} (y;\alpha )$ and $g_{\textrm{BN}} (y;\alpha )$ in (\ref{eq:Px_hat}),
it is straightforward to derive that
\begin{equation}
\label{eq40}
E_Y \{\hat {x}_{\textrm{BN}}^2 (y)\}=\sum\limits_{m=0}^\infty {\beta _m \sigma
_{y,m}^2 \left[ {\textrm{erf}\left( {\frac{\alpha }{\sqrt 2 \sigma _{y,m} }}
\right)-\sqrt {\frac{2}{\pi }} \frac{\alpha }{\sigma _{y,m}
}e^{-\frac{\alpha ^2}{2\sigma _{y,m}^2 }}} \right]}
\end{equation}
and
\begin{equation}
\label{eq41}
E_Y \{\hat {x}_{\textrm{SL}}^2 (y)\}=E_Y \{\hat {x}_{\textrm{BN}}^2 (y)\}+\alpha
^2\sum\limits_{m=0}^\infty {\beta _m \left[ {1-\textrm{erf}\left( {\frac{\alpha
}{\sqrt 2 \sigma _{y,m}}} \right)} \right]} .
\end{equation}
Note that (\ref{eq37}), (\ref{eq39}), (\ref{eq40}) and (\ref{eq41})
are different from the similar equations in \cite{Zhidkov:2006} and \cite{Zhidkov:2008}:
indeed, this paper deals with SL and BN of real random variables,
while \cite{Zhidkov:2006} and \cite{Zhidkov:2008} consider the SL and
the BN for the envelope of complex random variables.
Plugging (\ref{eq38}) and (\ref{eq40}) in (\ref{eq:x_SLBN_def_der}),
after some algebraic manipulation the optimal BN threshold
$\alpha _{\textrm{BN}}^{(snr)} $ is the solution of the
following equation
\begin{equation}
\label{eq42}
\frac{\sum\limits_{m=0}^\infty {\frac{\beta _m }{\sqrt 2 \sigma _{y,m}
}e^{-{\alpha ^2} \mathord{\left/ {\vphantom {{\alpha ^2} {2\sigma _{y,m}^2
}}} \right. \kern-\nulldelimiterspace} {2\sigma _{y,m}^2 }}} }{E\left\{
{\hat {x}_{\textrm{BN}}^2 (y)} \right\}}-\frac{\sum\limits_{m=0}^\infty {\frac{\sqrt
2 \beta _m }{\sigma _{y,m}^3 }e^{-{\alpha ^2} \mathord{\left/ {\vphantom
{{\alpha ^2} {2\sigma _{y,m}^2 }}} \right. \kern-\nulldelimiterspace}
{2\sigma _{y,m}^2 }}} }{k_x^{\textrm{(BN)}} }=0.
\end{equation}
Analogously, plugging (\ref{eq37}) and (\ref{eq41}) in (\ref{eq:x_SLBN_def_der}), the optimal SL threshold
$\alpha _{\textrm{SL}}^{(snr)} $ is the solution of the following equation
\begin{equation}
\label{eq43}
\frac{\sum\limits_{m=0}^\infty {\alpha \beta _m \left( {1-\textrm{erf}\left(
{\frac{\alpha }{\sqrt 2 \sigma _{y,m} }} \right)} \right)} }{E\left\{ {\hat
{x}_{\textrm{SL}}^2 (y)} \right\}}-\frac{\sqrt {\frac{2}{\pi }}
\sum\limits_{m=0}^\infty {\frac{\beta _m }{\sigma _{y,m} }e^{-{\alpha ^2}
\mathord{\left/ {\vphantom {{\alpha ^2} {2\sigma _{y,m}^2 }}} \right.
\kern-\nulldelimiterspace} {2\sigma _{y,m}^2 }}} }{k_x^{\textrm{(SL)}} }=0.
\end{equation}
Equations (\ref{eq42}) (\ref{eq43}) can be obviously solved by
root-finding numerical techniques \cite{Mayers:2003}, (\ref{eq43})
However it is also possible to cast them in a fixed-point problem,
which can be solved by iterative numerical approaches similar to A1 and A2.
For instance the equivalent formulation of (\ref{eq43}) is expressed by
\begin{equation}
\label{eq44}
\alpha =F_{\textrm{SL}}^{(snr)} (\alpha )=\frac{E\left\{ {\hat {x}_{\textrm{SL}}^2 (y)}
\right\}}{k_x^{\textrm{(SL)}} } \frac{\sqrt {\frac{2}{\pi }}
\sum\limits_{m=0}^\infty {\frac{\beta _m }{\sigma _{y,m} }e^{-{\alpha ^2}
\mathord{\left/ {\vphantom {{\alpha ^2} {2\sigma _{y,m}^2 }}} \right.
\kern-\nulldelimiterspace} {2\sigma _{y,m}^2 }}}
}{1-\sum\limits_{m=0}^\infty {\beta _m \textrm{erf}\left( {\frac{\alpha }{\sqrt 2
\sigma _{y,m} }} \right)} }.
\end{equation}
It is interesting to observe that (\ref{eq44}) can be rearranged as
\begin{equation}
\label{eq45}
\alpha =F_{\textrm{SL}}^{(snr)} (\alpha )=\frac{E\left\{ {\hat {x}_{\textrm{SL}}^2 (y)}
\right\}}{\sigma_X^2 k_x^{\textrm{(SL)}} } F_{\textrm{SL}}^{(mse)} (\alpha ),
\end{equation}
which means that the MSNR solution for the SL is different from the
MMSE solution in (\ref{eq:alfaSL_FixPoint_mse}). However, when the optimal thresholds
are sufficiently higher than the input standard deviation $\sigma _y $,
the power of the distortion noise is quite low with $E\{\hat {x}_{SL}^2 \}\approx k_x^2 \sigma_X^2 $,
which together with (\ref{eq45}) means that the two optimal thresholds
are very close if $k_x \approx 1$.
This specific observation for the SL can be generalized
by exploiting (\ref{eq:MSE_new_def}), which allows to
conclude that the MMSE thresholds are obtained by
\begin{equation}
\label{eq:alfa_XX_mmse_def}
\begin{array}{rcl}
\alpha _{\textrm{XX}}^{(mse)} &=&\mathop {\arg \min }\limits_{\alpha  \in {\cal R}^+} \left[ {E_e
\{e^2\}} \right] = \mathop {\arg \min }\limits_{\alpha  \in {\cal R}^+} \left[ {{E_Y \{\hat {x}_{\textrm{XX}}^2 (y)\}}
\mathord{\left/ {\vphantom {{E_Y \{\hat {x}_{\textrm{XX}}^2 (y)\}} {k_x^{\textrm{(XX)}} }}}
\right. \kern-\nulldelimiterspace} {k_x^{\textrm{(XX)}} }} \right] \\
&=&\mathop {\arg \min }\limits_{\alpha  \in {\cal R}^+  } \left[ {{{\left( {(k_x^{({\text{XX}})} )^2 \sigma _X^2  + \sigma _{W_x }^2 } \right)} \mathord{\left/ {\vphantom {{\left( {(k_x^{({\text{XX}})} )^2 \sigma _X^2  + \sigma _{W_x }^2 } \right)} {k_x^{({\text{XX}})} }}} \right.
 \kern-\nulldelimiterspace} {k_x^{({\text{XX}})} }}} \right].
\end{array}
\end{equation}
Thus, the MMSE criterion in (\ref{eq:alfa_XX_mmse_def}) is different from the MSNR criterion
in (\ref{eq:alfa_XX_msnr_def}) due to the absence of the square-power in the
denominator of the cost function.
Consequently, MMSE and MSNR approaches provide very close thresholds when
$k_x^{\textrm{(XX)}} \approx 1$: this happens for instance when the
clipping threshold $\alpha$ is sufficiently higher than $\sigma _y $,
due to $\sigma_N^2 \ll \sigma_X^2 $ [see also (\ref{eq:OBE_lin})].


\section{Theoretical MSE ans SNR computation}
\label{sec:SNR-MSE-theory}
According to (\ref{eq:e_SL}) and (\ref{eq:e_BN}) the MSE should be computed by
\begin{equation}
\label{eq:MSE_def}
E\{e^2\}=E\{h_{XX}^2(x,n;\alpha )\}=\sum_{m=0}^{\infty}{\beta_m E_{XN_m}\{h_{XX}^2(x,n_m;\alpha)\}},
\end{equation}
which requests tedious double-folded integrals with respect to the signal and the noise \emph{pdf}s.
However, exploiting (\ref{eq:x_hat_kX}), the estimation error can be also expressed by
\begin{equation}
\label{eq:e_new_def}
e=x-\hat {x}(y)=(1-k_x)x-w_x.
\end{equation}
Thus, due to the orthogonality of $x$ and $w_x$, an alternative expression for the MSE is
\begin{equation}
\label{eq:MSE_new_def}
\begin{array}{rcl}
E\{e^2\} &=& \left(1-k_x\right)^2\sigma_X^2 +E_{W_x } \{w_x^2 \}\\
&=&\left(1-2k_x\right)\sigma_X^2 +E_Y \{\hat {x}^2(y)\},
\end{array}
\end{equation}
where the last equality comes from $E_Y \{\hat{x}^2 (y)\}= k_x
^2\sigma_X^2 +E_{W_x } \{w_x^2 \}$.
This alternative expression
is very useful for the computation of the MSE of any non-linear
estimator because, differently from (\ref{eq:MSE_def}),
it requests to compute only single-folded integrals,
 e.g., the estimator average output power $E_Y \{\hat {x}_{
\textrm{XX}}^2 (y)\}$ by (\ref{eq:Px_hat}) and the gain $k_x^{\textrm{(XX)}}$ by (\ref{eq:kX_sum}).
Actually, for the suboptimal estimators considered in this paper,
these single-folded integrals are known in closed form for any $\alpha$,
as expressed by (\ref{eq40}) and (\ref{eq38})
for the BN, and  (\ref{eq41}) and (\ref{eq39}) for the SL.
Thus, plugging in these equations the values of $\alpha_{\textrm{(XX)}}^{(mse)}$ (or $\alpha_{\textrm{(XX)}}^{(snr)}$) obtained
by the MMSE (or MSNR) criterion allow to compute the corresponding
theoretical expressions for the MSE of the two suboptimal estimators.
The same considerations hold true for the theoretical SNR, whose analytical expression
in (\ref{eq:SNRout_def}) requires the computation of the same single-folded integrals used for the MSE.
Actually, exploiting the last equality in (\ref{eq:SNRout_def}) and plugging $E_Y \{\hat{x}(y)^2\}$ in (\ref{eq:MSE_new_def}),
it is derived that for any estimator $g(y)$ the link between MSE and SNR is expressed by
\begin{equation}
\label{eq:MSEvsSNR}
\textrm{MSE} = (k_x-1)^2 \sigma_X^2 - \frac{k_x^2\sigma_X^2}{\textrm{SNR}+1}.
\end{equation}


\section{Computer Simulations}
\label{sec:computer}
This section verifies by computer simulations the analytical results derived so far.
The MSE and SNR performance of the SL and BN are computed for several sets
of the Class-A canonical parameters $A$, $T$, $\sigma_N^2$
and for several SNR values. All the simulated MSEs and SNRs are obtained by
generating $10^9$ observed samples $y$ in (\ref{eq:sig-plus-noise}).
The Middleton's Class-A noise has been generated by the toolbox \cite{Interference:1}.
The optimal MMSE thresholds for the SLE and BNE are obtained by A1 and A2, using
$\varepsilon =0.01$ and $\mu =0.01$. The MSNR thresholds for SL and BN are obtained by
Matlab$^{\textrm{{\textregistered}}}$ numerical solutions of (\ref{eq43}) and (\ref{eq42}),
respectively. The series with infinite terms, which are induced in all the
analytical results by the Class-A \emph{pdf} in (\ref{eq:noise_pdf}),
have been approximated by considering only the first $M=50$ terms
(although, $M\in [10,20]$ would be enough in most of the cases).

\begin{figure}[h]
\centerline{\includegraphics[trim=0cm 0cm 0cm 0.7cm, clip=true, width=\figwidth]{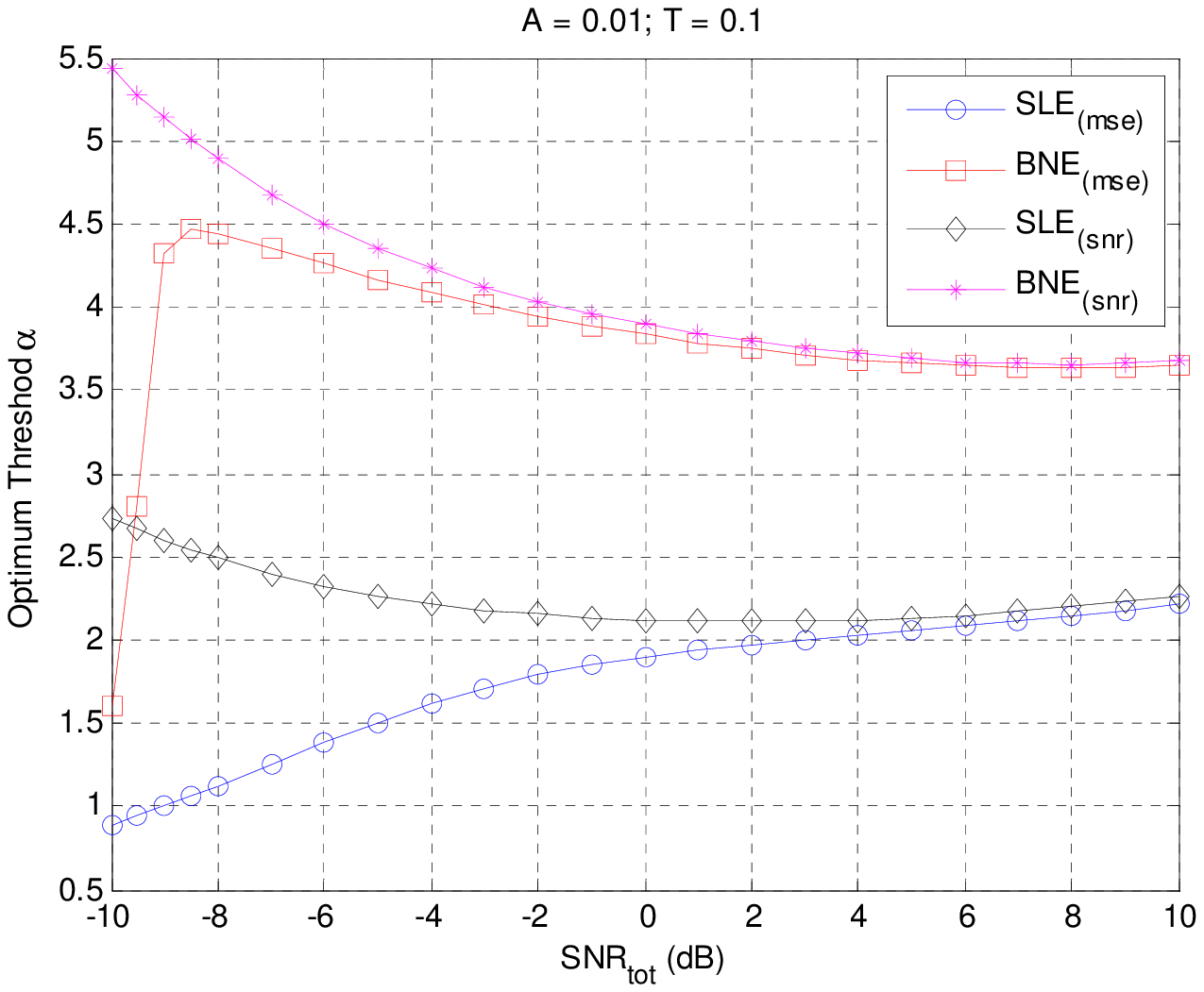}}
\caption{Optimal SL and BN thresholds ($A=0.01$, $T=0.1$, $\sigma_X^2 =1)$}
\label{fig6}
\end{figure}

\begin{figure}[h]
\centerline{\includegraphics[trim=0cm 0cm 0cm 0.7cm, clip=true, width=\figwidth]{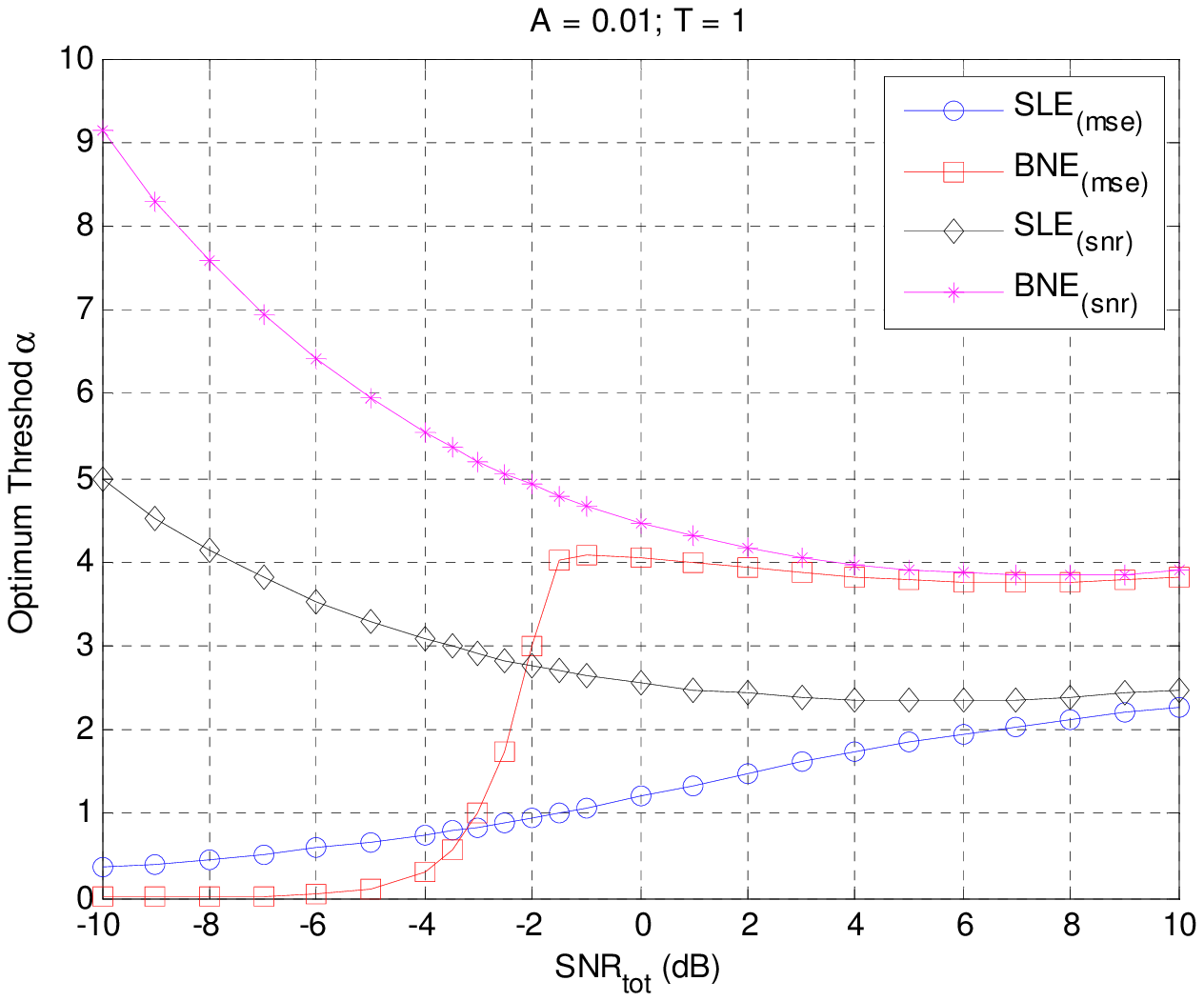}}
\caption{Optimal SL and BN thresholds ($A=0.01$, $T=1$, $\sigma_X^2 =1)$.}
\label{fig7}
\end{figure}

In \figurename~\ref{fig6} and \figurename~\ref{fig7}
it is possible to observe the dependence of the optimal SL and BN
thresholds on the total SNR, which is defined as $\textrm{SNR}_{\textrm{tot}} =\sigma_X^2
/(\sigma _t^2 +\sigma _I^2 )$. It is shown that the MMSE and MSNR (optimal)
thresholds are similar for high values of $\textrm{SNR}_{\textrm{tot}}$
(i.e., when $\sigma_X^2 \gg \sigma_N^2$) and consequently
the two criteria are almost equivalent.
This is not the case for low (and negative) values of the $\textrm{SNR}_{\textrm{tot}}$,
where the MMSE and MSNR thresholds tend to diverge.
Moreover, it is worth noting that the MMSE and MSNR thresholds are
more different for the BN rather than for the SL. This fact is more evident
when $T=\sigma _t^2 /\sigma _I^2 =1$, i.e., when the noise power is equally
split between the AWGN and the impulsive component.

\begin{figure}[h]
\centerline{\includegraphics[trim=0cm 0cm 0cm 0.7cm, clip=true, width=\figwidth]{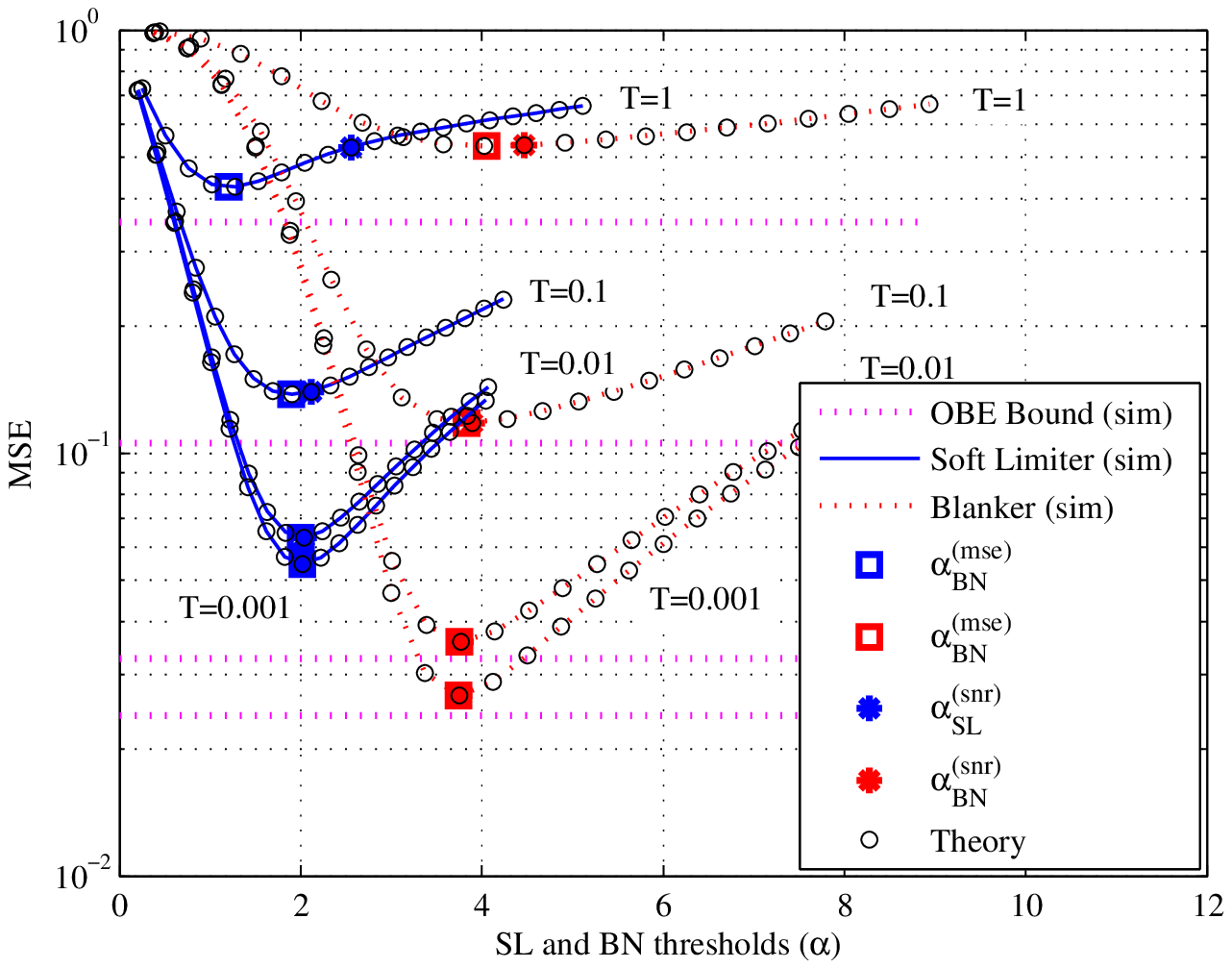}}
\caption{MSE curves for $A=0.01$ and $SNR_{tot} =0\textrm{ dB}$ }
\label{fig8}
\end{figure}

\figurename~\ref{fig8}-\figurename~\ref{fig13} let better
appreciate the sensitiveness of the SNR and MSE
performance with respect to the SL and BN thresholds,
as well as the performance penalties od the two
suboptimal estimators with respect to the OBE.
All the figures show that the minimum MSE and the maximum SNR
are obtained for the optimal thresholds values predicted by the theory.
Moreover, also the theoretical MSE and SNR derived in this paper
perfectly match with the simulation results.
As anticipated, in several scenarios the optimal MMSE and MSNR
thresholds are almost equivalent, and consequently they provide almost
the same MSE and SNR performance.
However, this is not the case in highly critical scenarios where
the $\textrm{SNR}_{\textrm{tot}}$ is quite low or negative.
This behavior is amplified when the AWGN noise power
is not negligible with respect to the impulsive noise power
(i.e., $T\approx 1$ ) or when also the impulsive noise tends
to be Gaussian (i.e., $A\approx 1$)). As theoretically expected,
all the figures also highlight that the OBE always
outperforms in MSE the SLE and the BNE. However, the SLE and BNE
penalties are not dramatic, as it was expected by the fact that the OBE
shapes in \figurename~\ref{fig2} highly resemble either the SLE or the BNE
for several values of the canonical Class-A parameters. Interestingly,
although the OBE is not the MSNR optimal estimator,
in most of the cases it outperforms in SNR the MSNR-optimal BN and SL.
As a final remark, the theoretical results shown in this paper
can be directly employed to predict the MSE and SNR performance
of multicarrier telecommunication systems (such as ADSL and PLC)
that employ the proposed estimators to contrast an impulsive
interference modeled as a Gaussian-mixture \cite{Nassar:2011}.

\section*{CONCLUSIONS}
This paper has derived the MMSE Bayesian estimator for a Gaussian source
impaired by impulsive Middleton's Class-A interference. The estimator is
directly extensible to any Gaussian-mixture noise. Two popular and
sub-optimal estimators, namely the soft-limiter and the blanker, have been
optimized both in a MSE and SNR sense, deriving also closed form expressions
for their MSE and SNR. A theoretical link between MSE and SNR at the output
of the estimator has been established, and scenarios when the MMSE and the
maximum SNR criteria are (almost) equivalent, or different, have been
clarified. The theoretical analysis and computer simulations have shown that
at least one estimator, among the optimum soft-limiter or the optimum
blanker, can be always used as a sub-optimum estimator with minimal
performance loss with respect to the MMSE Bayesian estimator.
The derivation of the optimal estimator in the maximum-SNR sense, as well as
a closed-form expression for the MSE of the optimal Bayesian estimator,
are still open problems for possible further research.


\begin{figure}[h]
\centerline{\includegraphics[trim=0cm 0cm 0cm 0.7cm, clip=true, width=\figwidth]{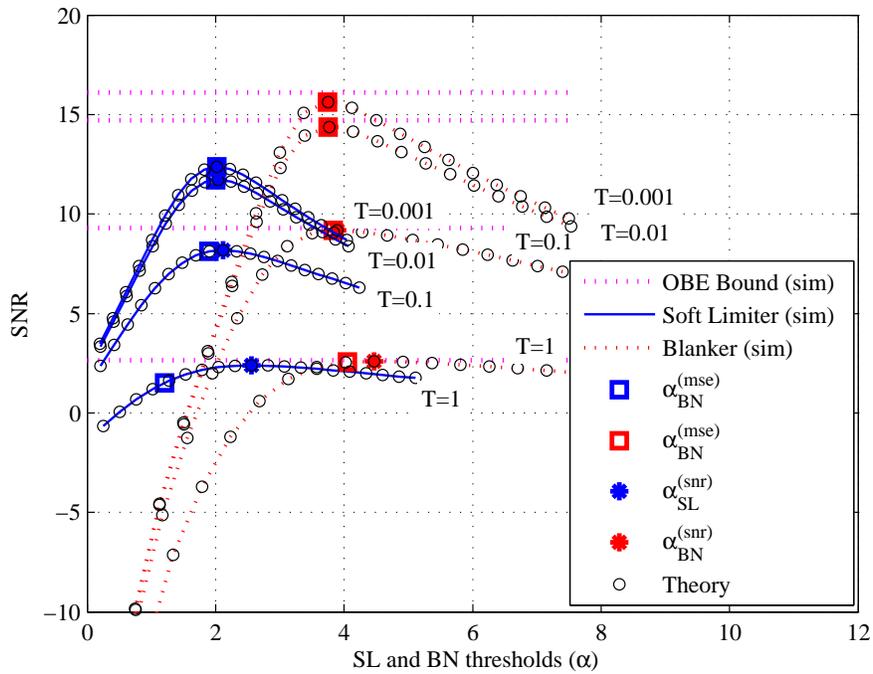}}
\caption{SNR curves for $A=0.01$ and $SNR_{tot} =0\textrm{ dB}$ }
\label{fig9}
\end{figure}

\begin{figure}[h]
\centerline{\includegraphics[trim=0cm 0cm 0cm 0.7cm, clip=true, width=\figwidth]{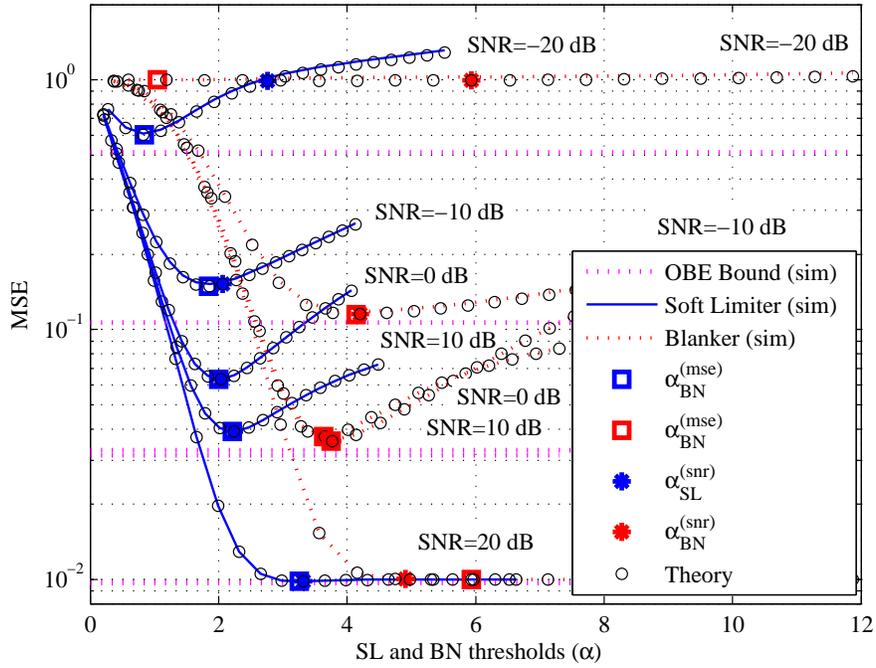}}
\caption{MSE curves for $A=0.01$ and $T=0.01\textrm{ }$ }
\label{fig10}
\end{figure}

\begin{figure}[h]
\centerline{\includegraphics[trim=0cm 0cm 0cm 0.7cm, clip=true, width=\figwidth]{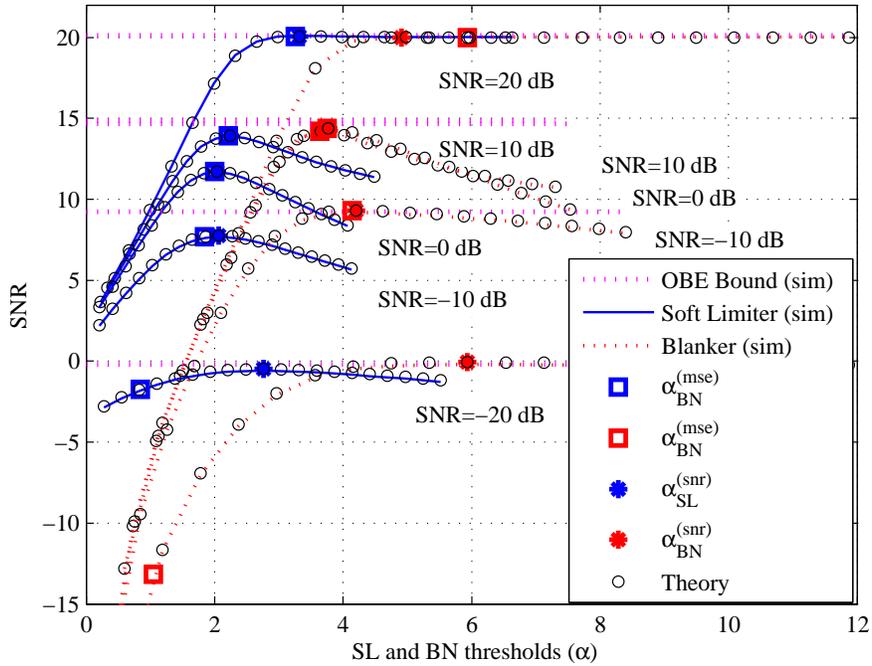}}

\caption{SNR curves for $A=0.01$ and $T=0.01\textrm{ }$}
\label{fig11}
\end{figure}

\begin{figure}[h]
\centerline{\includegraphics[trim=0cm 0cm 0cm 0.7cm, clip=true, width=\figwidth]{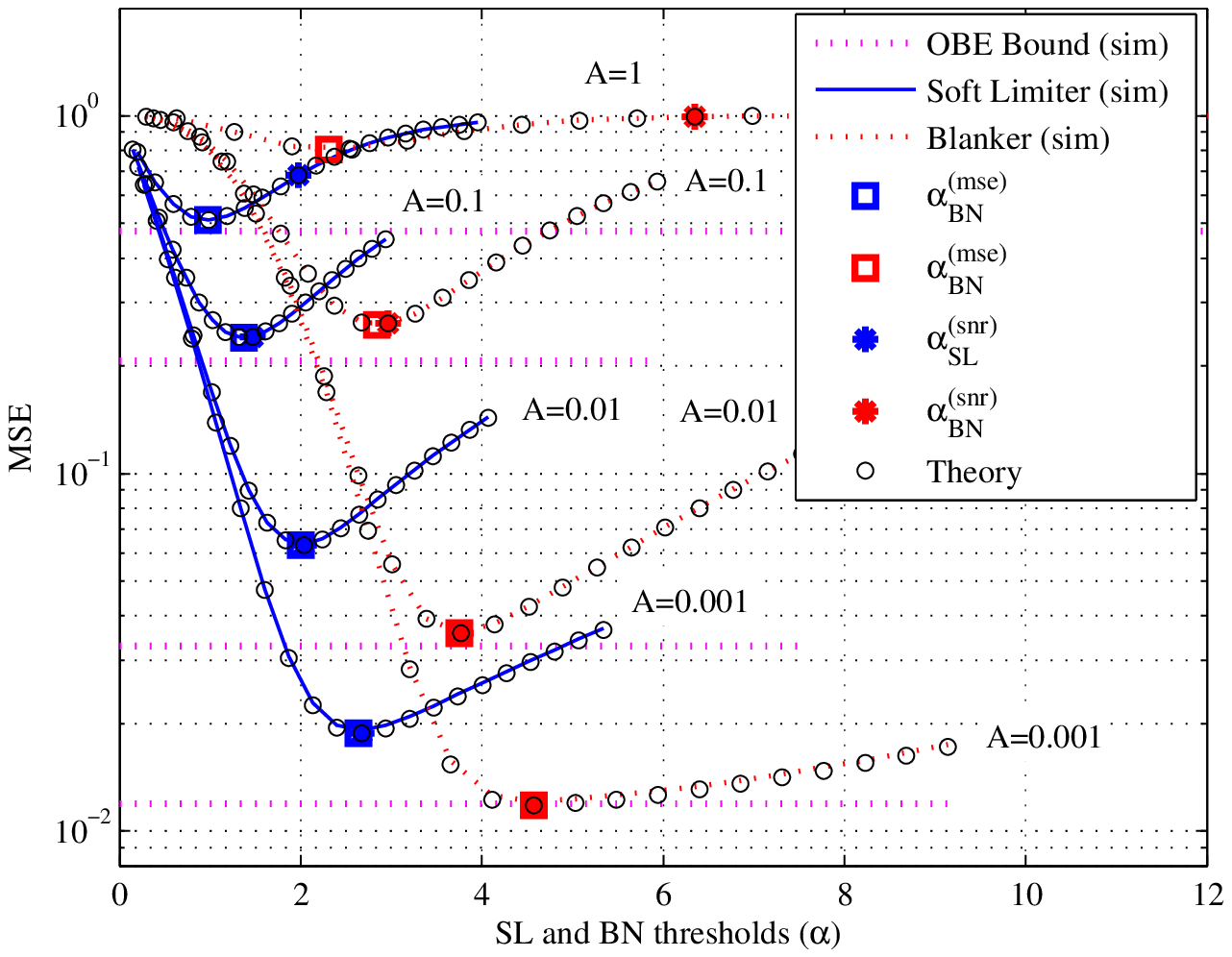}}

\caption{MSE curves for $T=0.001$ and $SNR_{tot} =0\textrm{ dB}$}
\label{fig12}
\end{figure}

\begin{figure}[h]
\centerline{\includegraphics[trim=0cm 0cm 0cm 0.7cm, clip=true, width=\figwidth]{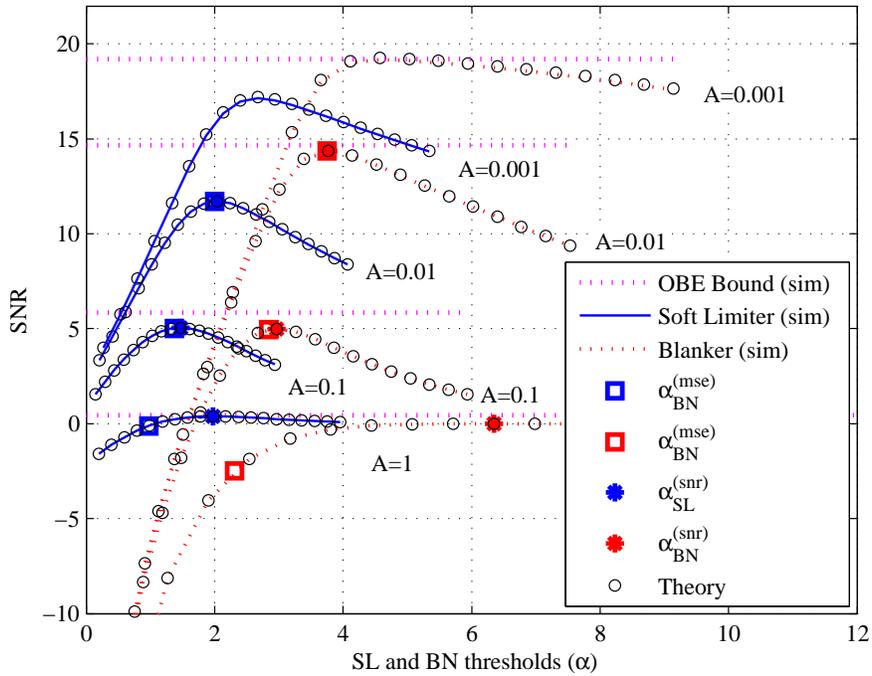}}
\caption{SNR curves for $T=0.001$ and $SNR_{tot} =0\textrm{ dB}$}
\label{fig13}
\end{figure}

%

\begin{figure}[h]
\centerline{\includegraphics[trim=0cm 0cm 8cm 19cm, clip=true, width=0.7\figwidth]{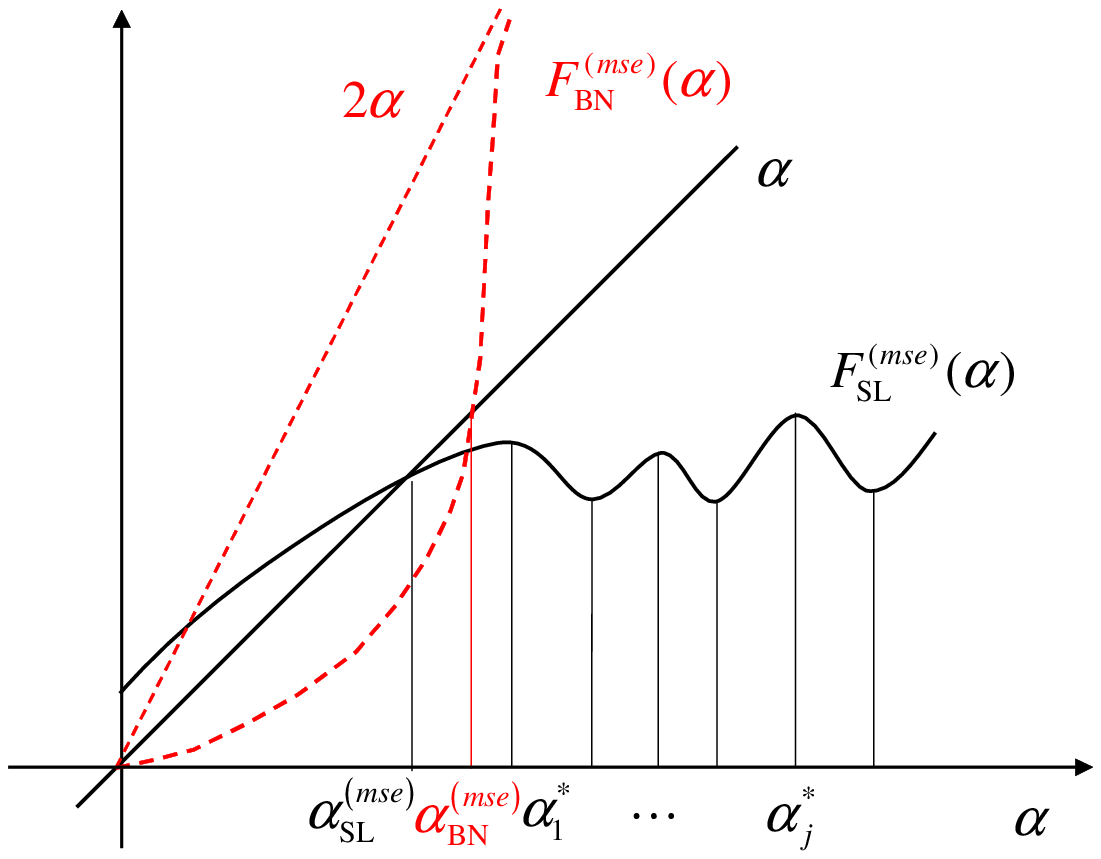}}
\caption{Typical fixed-point problems for the SLE and the BNE}
\label{fig14}
\end{figure}

\appendices


\section{OBE derivation}
\label{App:OBE}
$S(f)=\mathpzc{F}\left\{ {s(x)} \right\}=\int\limits_{-\infty }^{+\infty }
{s(x)e^{-j2\pi fx}dx}$ is used to indicate the FT
of $s(x)$.
It is also reminded that the FT of a Gaussian \emph{pdf} is still a normalized Gaussian function,
expressed by $\tilde {G}(f;\sigma _{X,f}^2 )=\sqrt{2 \pi\sigma_{X,f}^2} G(f;\sigma_{X,f}^2)$,
where $\sigma _{X,f}^2 =1/(4\pi \sigma_X^2 )$. Thus, by exploiting the convolution and derivative
properties of the FT, the integral in (\ref{eq:OBE_conv}) is expressed in the frequency domain by
\begin{equation}
\label{eq53}
\begin{array}{rcl}
\mathpzc{F}\left\{ {p_x (x)\ast f_m (x)} \right\}
&=&\mathpzc{F}\left\{ xG(x;\sigma_X^2) \right\}\mathpzc{F}\left\{ {G(x;\sigma _m^2 )}
\right\}
\\
 &=& \frac{j}{2\pi }\frac{d}{df}\left[ {\tilde {G}(f;\sigma _{X,f}^2 )}
\right]\tilde {G}(f;\sigma_{m,f}^2 ). 
 \end{array}
\end{equation}
By observing that $\frac{d}{df}\left[ {G(f;\sigma^2 )} \right]=-\frac{f}{\sigma^2}G(f;\sigma^2)$,
then (\ref{eq53}) becomes
\begin{equation}
\label{eq55}
\mathpzc{F}\left\{ {p_x (x)\ast f_m (x)} \right\}
=-\frac{j}{2\pi \sigma _{X,f}^2}f\textrm{ }\tilde {G}(f;\sigma _{X,f}^2 )\tilde {G}(f;\sigma _{m,f}^2 ).
\end{equation}
Thus, by the duality property of the inverse FT
\begin{equation}
\label{eq:OBEconv_by_FT}
\begin{array}{rcl}
 p_x (x)\ast f_m (x)
&=&-\frac{1}{4\pi ^2\sigma _{X,f}^2 }\frac{d}{dx}\left[ {\mathpzc{F}^{-1}\left\{
{\tilde {G}(f;\sigma _{X,f}^2 )\tilde {G}(f;\sigma _{m,f}^2 )} \right\}}
\right] \\
&=&-\sigma_X^2 \frac{d}{dx}\left[ {G(x;\sigma_X^2 )\ast G(x;\sigma _m^2 )}
\right]. \\
 \end{array}
\end{equation}
The convolution of two zero-mean Gaussian $pdfs$ is still a zero-mean Gaussian \emph{pdf}
with a variance equal to the sum of the two single variances, and consequently
(\ref{eq:OBEconv_by_FT}) becomes
\begin{equation}
\label{eq56}
p_x(x)\ast f_m (x)
=\frac{\sigma_X^2 }{\sigma_X^2 +\sigma _m^2 }xG(x;\sigma_X^2 +\sigma_m^2 ).
\end{equation}
Summarizing, equation (\ref{eq:OBE_conv}) can be expressed by
\begin{equation}
\label{eq57}
\begin{array}{rcl}
\hat {x}_{\textrm{OBE}} (y)
&=&\frac{1}{f_Y (y)}\sum\limits_{m=0}^\infty {\beta _m
\left[ {p_x(y)\ast f_m (y)} \right]}
\\
&=&\frac{\sum\limits_{m=0}^\infty
{\frac{\sigma_X^2 \beta _m }{\sigma_X^2 +\sigma _m^2 }G(y;\sigma_X^2
+\sigma _m^2 )} }{\sum\limits_{m=0}^\infty {\beta _m G(y;\sigma_X^2 +\sigma
_m^2 )} }y,
\end{array}
\end{equation}
which coincides with (\ref{eq:OBE_final}).

\section{SLE derivation}
\label{App:SLE}
By observing (\ref{eq:e_SL}) it is clear that $h_{\textrm{SL}} (x,n;\alpha )$ is continuous with
respect to $\alpha \in \cal {R}^+$ and
\begin{equation}
\label{eq:hSL_der_hSL}
h_{\textrm{SL}} (x,n;\alpha )h_{\textrm{SL}}^{(1,\alpha )} (x,n;\alpha )=\left\{
{\begin{array}{ll}
 x+\alpha 	\quad &,x<-\alpha -n \\
 0	\quad &,\vert x+n\vert \le \alpha \\
 \alpha -x	\quad &,x>-n+\alpha \\
 \end{array}} \right.\quad .
\end{equation}
Consequently, substituting (\ref{eq:hSL_der_hSL}) in (\ref{eq:der_alfaSL}), the optimum value $\alpha_{SLE}^{(mse)}$
is a solution of
\begin{equation}
\label{eq:J_SL_derivative}
\begin{array}{rcl}
J_{\textrm{SL}}^{(1,\alpha)}(\alpha)
&=&\int\limits_{-\infty }^{+\infty } {\int\limits_{-\infty }^{-\alpha -n}
{(x+\alpha )f_X (x)dx} f_N (n)dn} \\
&\qquad& \quad +\int\limits_{-\infty }^{+\infty }
{\int\limits_{\alpha -n}^{+\infty } {(\alpha -x)f_X (x)dx} f_N (n)dn} =0,
\end{array}
\end{equation}
which by standard integration of Gaussian density functions is equivalent to
\begin{equation}
\label{eq:alfa_Nsl_by_Dsl}
\alpha =\frac{N_{\textrm{SL}} (\alpha )}{D_{\textrm{SL}} (\alpha )}
=\frac{\frac{\sigma_X}{\sqrt {2\pi}}
\int\limits_{-\infty }^{+\infty}
{\left[
e^{-\frac{(\alpha+n)^2}{2\sigma_X^2}}
+ e^{-\frac{(\alpha-n)^2}{2\sigma_X^2}}\right] f_N (n)dn}}
{\int\limits_{-\infty }^{+\infty } {\left[ {\Phi \left( {-\frac{n+\alpha
}{\sigma_X }} \right)+\Phi \left( {\frac{n-\alpha }{\sigma_X }} \right)}
\right]f_N (n)dn} },
\end{equation}
where
\begin{equation}
\label{eq:J_SL_by_Nsl_Dsl}
J_{\textrm{SL}}^{(1,\alpha)}(\alpha)= \alpha D_{\textrm{SL}}(\alpha) - N_{\textrm{SL}}(\alpha)
\end{equation}
and $\Phi (x)=0.5\left[1+\textrm{erf}\left(x/ \sqrt{2}\right)\right]$.
By substituting in (\ref{eq:alfa_Nsl_by_Dsl}) the expression of $f_N (n)$ in (\ref{eq:noise_pdf}), it is
recognized that the numerator
\begin{equation}
\label{eq:Nsl_integral}
\begin{array}{l}
N_{\textrm{SL}} (\alpha )=\sigma_X^2 \sum\limits_{m=0}^\infty
{\frac{\beta_m}{2}\left[ {\int\limits_{-\infty }^{+\infty } {G(n;\sigma _m^2
)G(\alpha -n;\sigma_X^2 )dn} } \right.} \\
\left.
\qquad \qquad \qquad \qquad \qquad
 {+\int\limits_{-\infty
}^{+\infty } {G(n;\sigma _m^2 )G(\alpha +n;\sigma_X^2 )dn} } \right]
\end{array}
\end{equation}
contains two integrals representing the convolution, and the correlation,
of two zero-mean Gaussian $pdfs$.
Due to the even symmetry of Gaussian functions, the correlation
is equivalent to the convolution and the result is another
zero-mean Gaussian \emph{pdf}, as expressed by
\begin{equation}
\label{eq:Nsl_final}
\begin{array}{rcl}
 N_{\textrm{SL}}(\alpha)&=&2\sigma_X^2 \sum\limits_{m=0}^\infty {\beta_m G(\alpha;\sigma_X^2 + \sigma_m^2)}\\
 &=&2\sigma_X^2\sum\limits_{m=0}^\infty{
\beta_m \frac{e^{-\alpha^2/2(\sigma_X^2 +\sigma_m^2)}}{\sqrt{2\pi(\sigma_X^2 +\sigma_m^2)}}}
 =2\sigma_X^2 \sum \limits_{m=0}^\infty{N_m(\alpha)}.
 \end{array}
\end{equation}
By observing that $G(n;\sigma^2)=G(-n;\sigma^2)$ it is possible to
recognize that
\begin{equation}
\label{eq:Dsl_integral}
\begin{array}{rcl}
D_{\textrm{SL}} (\alpha )&=&\sum\limits_{m=0}^\infty{
\beta_m \int\limits_{-\infty }^{+\infty } {\left[ {\Phi \left(
{\textstyle{{-n-\alpha } \over {\sigma_X }}} \right)+\Phi \left(
{\textstyle{{n-\alpha } \over {\sigma_X }}} \right)} \right]G(n;\sigma _m^2
)dn} }
\\
&=& \sum\limits_{m=0}^\infty{\beta_m \left[ q(-\alpha)+q(\alpha)\right]},
\end{array}
\end{equation}
where the convolution $q(\alpha)= \Phi(\alpha/\sigma_X) \ast G(\alpha;\sigma_m^2)$
can be easily solved in the FT domain.
Indeed, exploiting the integral property of the FT
$\mathpzc{F}\{\Phi(\alpha/\sigma_X)\}=\left[\frac{1}{j2 \pi f}
+ \frac{1}{2}\delta(f)\right] \mathpzc{F}\{G({\alpha;\sigma_X^2})\}$
it follows that
\begin{equation}
\label{eq:q_alpha}
\begin{array}{rcl}
q(\alpha)&=&\mathpzc{F}^{-1}\left\{\left[\frac{1}{j2 \pi f} + \frac{1}{2}\delta(f)\right] \tilde{G}\left(f;\sigma_{f,X}^2\right)\tilde{G}\left(f;\sigma_{f,m}^2\right)\right\}
\\
&=&\mathpzc{F}^{-1}\left\{\left[\frac{1}{j2 \pi f} + \frac{1}{2}\delta(f)\right] \tilde{G}\left({f;(\sigma_{f,X}^{-2}+\sigma_{f,m}^{-2})^{-1}}\right)\right\}
\\
&=& \int\limits_{-\infty}^{\alpha}{G\left({z;\sigma_X^2+\sigma_m^2}\right)dz} = \Phi\left(\alpha/\sqrt{\sigma_X^2+\sigma_m^2}\right),
\end{array}
\end{equation}
which together with $\Phi(x)=1-\Phi(-x)$ lets to conclude
\begin{equation}
 D_{\textrm{SL}} (\alpha ) =
 1-\sum\limits_{m=0}^\infty{\beta_m \textrm{erf}\left( \frac{\alpha}{\sqrt{2(\sigma _m^2 +\sigma_X^2)} }\right)}.
\end{equation}
To prove the existence of a solution of the fixed point equation (\ref{eq:alfaSL_FixPoint_mse}),
it can be observed that the relative minima and maxima of $F_{\textrm{SL}}^{(mse)}
(\alpha )$ are obtained by the zeros of
\begin{equation}
\label{eq64}
\begin{array}{l}
 F_{\textrm{SL}}^{(1,\alpha )} (\alpha )=\frac{2\sigma_X^2 }{D_{\textrm{SL}} (\alpha
)^2}\left[ {-\sum\limits_{m=0}^\infty {\frac{\alpha }{\sigma_X^2 +\sigma_m^2 }N_m (\alpha )D_{\textrm{SL}} (\alpha )} } \right. \\
\qquad \qquad \qquad \qquad \qquad
+\left.{\sum\limits_{m=0}^\infty {\frac{2\beta _m }{\sqrt {2\pi (\sigma_X^2
+\sigma _m^2 )} }N_m (a)N_{\textrm{SL}} (\alpha )} } \right] \\
\qquad \qquad = 4\sigma_X^2 \frac{N_{\textrm{SL}}^2(\alpha)}{D_{\textrm{SL}}^2(\alpha )}-\frac{2\alpha
}{D_{\textrm{SL}} (\alpha )}\sum\limits_{m=0}^\infty {\frac{\sigma_X^2 }{\sigma_X^2
+\sigma _m^2 }N_m (a)} .
 \end{array}
\end{equation}
Thus, the relative minima and maxima correspond to values $\alpha ^\ast$ that satisfy the following expression
\begin{equation}
\label{eq65}
\begin{array}{rcl}
\alpha ^\ast &=&\frac{2\sigma_X^2 }{\sum\limits_{m=0}^\infty {\frac{\sigma_X^2 }{\sigma_X^2 +\sigma _m^2 }N_m (\alpha ^\ast )} }\frac{N_{\textrm{SL}}^2
(\alpha ^\ast )}{D_{\textrm{SL}} (\alpha ^\ast )}\\
&=&F_{\textrm{SL}}^{(mse)} (\alpha ^\ast
)\frac{\sum\limits_{m=0}^\infty {N_m (\alpha ^\ast )}
}{\sum\limits_{m=0}^\infty {\frac{\sigma_X^2 }{\sigma_X^2 +\sigma _m^2
}N_m (\alpha ^\ast )} }>F_{\textrm{SL}}^{(mse)} (\alpha ^\ast ) .
\end{array}
\end{equation}
Taking in mind that $F_{\textrm{SL}}^{(mse)} (0)=N_{\textrm{SL}}(0)>0$ and $F_{\textrm{SL}}^{(1,\alpha )}
(0)=4\sigma_X^2 N_{\textrm{SL}}(0)^2>0$, the inequality in (\ref{eq65}) means that all the relative
maxima and minima of $F_{\textrm{SL}}^{(mse)} (\alpha )$ occur when $F_{\textrm{SL}}^{(mse)} (\alpha )$ is
below the angle bisector $\alpha $, as shown in \figurename~\ref{fig14}.
Thus, $F_{\textrm{SL}}^{(mse)} (\alpha )$ should necessarily cross the angle bisector before its first relative maximum,
and consequently the fixed-point problem admits (at least) a solution.
The uniqueness of the fixed-point solution (and convergence of algorithm A1 to this solution)
would be granted if $F_{\textrm{SL}}^{(mse)}(\alpha )$ is a contraction mapping \cite{Mayers:2003}
between the fixed point solution and its first relative maximum (e.g., $\vert F_{\textrm{SL}}^{1,\alpha}
(\alpha )\vert <1$ for $\alpha _{opt} <\alpha <\alpha _1^\ast )$, as illustrated for more clarity in \figurename~\ref{fig13}.
Actually, this is equivalent to prove that the MSE minimization problem in (\ref{eq:alphaSL_min}) is locally convex, i.e.,
$\frac{\partial^2 }{\partial \alpha^2 }E\left\{ {h_{\textrm{SL}}^2 (x,n;\alpha )}\right\}>0$ for $\alpha \in [\alpha _{opt},\alpha_1^\ast]$.
To this end it is useful to express
\begin{equation}
\label{eq:der2_MSE_SL_def}
\begin{array}{rcl}
\frac{\partial^2}{\partial {\alpha}^2} E\left\{ {h_{\textrm{SL}}^2(x,n;\alpha )}\right\}
&=& 2E\left\{\left[h_{\textrm{SL}}^{(1,\alpha)}(x,n;\alpha)\right]^2\right\}
\\
&+& 2E\left\{h_{\textrm{SL}}(x,n;\alpha)h_{\textrm{SL}}^{(2,\alpha)}(x,n;\alpha)\right\}.
\end{array}
\end{equation}
Omitting detailed derivations, we simply observe that substituting
$h_{\textrm{SL}}^{(1,\alpha)}(x,n;\alpha)=-\textrm{sign}(x+n)u_{-1}(\vert x+n \vert-\alpha)$ and
$h_{\textrm{SL}}^{(2,\alpha)}(x,n;\alpha) = \textrm{sign}(x+n)\delta\left(\alpha-\vert x+n \vert \right)$,
the integrals in (\ref{eq:der2_MSE_SL_def}) can be solved exploiting (\ref{eq:q_alpha}) and (\ref{eq:Nbn_integral}),
to obtain
\begin{equation}
\label{eq:der2_MSE_SL}
\begin{array}{rcl}
\frac{\partial^2}{\partial {\alpha}^2} E\left\{ {h_{\textrm{SL}}^2(x,n;\alpha )}\right\}
&=& 2\left[ 1-\sum\limits_{m=0}^\infty{\beta_m\left(\Phi\left(\frac{\alpha}{\sqrt{\sigma_X^2+\sigma_m^2}}
\right)\right.} \right.
\\
&\qquad& + \left. \left. \frac{\sigma_m^2}{\sigma_X^2+\sigma_m^2}\frac{\alpha}{\sqrt{2\pi(\sigma_X^2+\sigma_m^2)}}
e^{-\frac{\alpha^2}{2(\sigma_X^2+\sigma_m^2)}} \right) \right].
\end{array}
\end{equation}
Although it is not easy to analytically prove that
$\frac{\partial^2 }{\partial \alpha^2 }E\left\{ {h_{\textrm{SL}}^2 (x,n;\alpha )}\right\}>0$
for any $\alpha \in [\alpha_{opt},\alpha_1^\ast]$, it can be observed that
surely $\frac{\partial^2 }{\partial \alpha^2 }E\left\{ {h_{\textrm{SL}}^2 (x,n;\alpha )}\right\} > 0$
when $\beta_m$ in (\ref{eq:der2_MSE_SL}) is multiplied by a coefficient lower than one for any $m$.
Thus, by noting that $(x/\sigma) e^{-x^2/2\sigma^2}\leq \/\sqrt{2\pi e}$, $\beta_m$
is always multiplied by a quantity lower than one when
 \begin{equation}
 \label{eq:alfa_SL_convex}
 \begin{array}{rcl}
\alpha &\leq&{\sqrt{\sigma_x^2+\sigma_m^2}}\Phi^{-1}\left(1-\frac{1}{\sqrt{2\pi e}}\frac{\sigma_m^2}{\sigma_X^2+\sigma_m^2}\right)
\\
 &<& 2.05 \sqrt{\sigma_x^2+\sigma_m^2}< 2.05 \sigma_x.
 \end{array}
 \end{equation}
This means that the algorithm A1 will converge toward the first minimum of the objective function in (\ref{eq:alphaSL_min})
every time is started with $\alpha_0 \in [0,2.05 \sigma_x]$,
Taking in mind that we are looking for the best soft-limiter threshold,
it is reasonable to infer that the above first minimum hit by the iterative algorithm
is also the optimal solution we are looking for, as also confirmed by the
simulation results.


\section{BNE derivation}
\label{App:BNE}
By observing (\ref{eq:e_BN}), it is possible to proceed with the same approach of
Appendix \ref{App:SLE}. Due to the fact that (\ref{eq:e_BN}) is a piecewise constant function with
respect to $\alpha  \in {\cal R}^+$ with a discontinuity in $\alpha =\vert x+n\vert
$, the Dirac's impulse function $\delta (x)$ can be exploited to handle the
derivative of $h_{\textrm{BN}}(x,n;\alpha )$ in this point. Thus, we obtain
\begin{equation}
\label{eq66}
\begin{array}{rcl}
 h_{\textrm{BN}}^{(1,\alpha )}(x,n;\alpha )&=&\frac{\partial h_{\textrm{BN}}(x,n;\alpha )}{\partial \alpha
} \\
 &=&[h_{\vert n+x\vert ^+} (x,n)-h_{\vert n+x\vert ^-} (x,n)]\delta (\alpha
-\vert n+x\vert ) \\
 &=&-(n+x)\delta (\alpha -\vert n+x\vert ), \\
 \end{array}
\end{equation}
where $h_{a^\pm } (x,n)=\mathop {\lim }\limits_{\alpha \to a^\pm }
h_{BN}(x,n;\alpha )$ represents the limit from either the right ($+$) or the left
($-$). Consequently, by direct substitution of (\ref{eq66})
\[
\begin{array}{lll}
&h_\alpha (x,n;\alpha )h^{(1,\alpha )}(x,n;\alpha )
\\
&\quad \quad=-(n+x)\frac{h_{\vert n+x\vert ^+} (x,n)+h_{\vert n+x\vert ^-}
(x,n)}{2}\delta (\alpha -\vert n+x\vert ) \\
&\quad \quad =\frac{(n^2-x^2)}{2}\delta (\alpha -\vert n+x\vert ).
 \end{array}
\]
Using for the BN the equivalent expression of (\ref{eq:SL_integral}), the
optimum $\alpha $ in the MMSE sense is obtained by equating to zero the derivative of the
MSE $J_{\textrm{BN}}(\alpha)$, as expressed by
\begin{equation}
\begin{array}{l}
J_{\textrm{BN}}^{(1,\alpha)}(\alpha)=\int\limits_{-\infty }^{+\infty }{\int\limits_{-\infty }^{-n}{(n^2-x^2)f_X
(x)} f_N (n)\delta (\alpha +x+n)dxdn} \\
\qquad +\int\limits_{-\infty }^{+\infty }
{\int\limits_{-n}^{+\infty } {(n^2-x^2)f_X (x)} f_N (n)\delta (\alpha
-x-n)dxdn}=0.
\end{array}
\end{equation}
By exploiting the integral properties of the Dirac's delta function, it is
possible to recognize that
\begin{equation}
\label{eq:BN_MSE_derivative_integral}
\begin{array}{rcl}
J_{\textrm{BN}}^{(1,\alpha)}(\alpha)&=&\int\limits_{-\infty }^{+\infty } {[n^2-(-n-\alpha )^2]f_X (-n-\alpha )f_N
(n)dn}\\
 &\quad& \quad + \int\limits_{-\infty }^{+\infty } {[n^2-(\alpha -n)^2]f_X (\alpha
-n)f_N (n)dn}.
\end{array}
\end{equation}
Exploiting the even symmetry of $f_X(x)$, equation (\ref{eq:BN_MSE_derivative_integral})
can be further simplified to
\begin{equation}
\label{eq67}
\begin{array}{l}
J_{\textrm{BN}}^{(1,\alpha)}(\alpha) = \alpha \int\limits_{-\infty }^{+\infty } {nf_N (n)[f_X (\alpha -n)-f_X(\alpha
+n)]dn}
\\
\qquad \qquad \qquad -\alpha^2 \int\limits_{-\infty }^{+\infty } {f_N (n)f_X (n+\alpha )dn} \\
\quad = \alpha\sum\limits_{m=0}^\infty {\beta _m \int\limits_{-\infty
}^{+\infty } {nf_m (n)[f_X (\alpha -n)-f_X (\alpha +n)]} dn}
\\
\qquad \qquad \qquad -\alpha^2\sum\limits_{m=0}^\infty {\beta _m \int\limits_{-\infty }^{+\infty } {f_m
(n)f_X (n+\alpha )dn}},
\end{array}
\end{equation}
where the first integral can be split as the difference of a
convolution and a correlation integral. Defining for convenience the odd function $g_m (n)=nf_m (n)$,
it can be observed that the first integral (\ref{eq67}) is expressed by
\begin{equation}
\label{eq:Nbn_integral}
\begin{array}{l}
 \int\limits_{-\infty }^{+\infty } {nf_m (n)[f_X (\alpha
-n)-f_X (\alpha +n)]dn}=
\\
\qquad \qquad 2g_m (\alpha )\ast f_X (\alpha )
    =2\mathpzc{F}^{-1}\{G_m (f)F_X (f)\}.
\end{array}
\end{equation}
Using the same approach that lead to (\ref{eq56}) in Appendix \ref{App:OBE},
the integral in (\ref{eq:Nbn_integral}) becomes
\begin{equation}
\begin{array}{l}
\int\limits_{-\infty }^{+\infty } {nf_m (n)[f_X (\alpha
-n)-f_X (\alpha +n)]dn} =
\\ \qquad \quad
\frac{2\sigma _m^2 }{\sqrt {2\pi } (\sigma _m^2 +\sigma_X^2 )^{3/2}}\alpha e^{-\alpha ^2/2(\sigma _m^2 +\sigma_X^2 )}.
\end{array}
\end{equation}
Thus, observing that the second integral in (\ref{eq67}) is just the convolution of two Gaussian
zero-mean \emph{pdf}s, the expression of the overall MSE derivative becomes
\begin{equation}
\label{eq:BN_MSE_derivative}
\begin{array}{ll}
J_{\textrm{BN}}^{(1,\alpha)}(\alpha) = \alpha \sum\limits_{m=0}^\infty {\frac{2\sigma _m^2
\beta _m }{(\sigma _m^2 +\sigma_X^2 )^{3/2}}\alpha e^{-\alpha ^2/2(\sigma
_m^2 +\sigma_X^2 )}}
\\
\qquad \qquad \quad
-\alpha^2 \sum\limits_{m=0}^\infty {\frac{\beta _m }{(\sigma
_m^2 +\sigma_X^2 )^{1/2}}e^{-\alpha ^2/2(\sigma _m^2 +\sigma_X^2 )}}.
\end{array}
\end{equation}
Equating to zero (\ref{eq:BN_MSE_derivative}) is equivalent to find the solution
of the fixed-point equation expressed by
\begin{equation}
\label{eq71}
\alpha =F_{\textrm{BN}}^{(mse)} (\alpha )=\frac{\sum\limits_{m=0}^\infty {\frac{2\sigma _m^2
\beta _m }{(\sigma _m^2 +\sigma_X^2 )^{3/2}}\alpha e^{-\alpha ^2/2(\sigma
_m^2 +\sigma_X^2 )}} }{\sum\limits_{m=0}^\infty {\frac{\beta _m }{(\sigma
_m^2 +\sigma_X^2 )^{1/2}}e^{-\alpha ^2/2(\sigma _m^2 +\sigma_X^2 )}} },
\end{equation}
which coincides with (\ref{eq:alfaBN_FixPoint_mse}).
The trivial solution $\alpha=0$ for (\ref{eq71}) and $J_{\textrm{BN}}^{(1,\alpha)}(\alpha) = 0$
corresponds to a BN output equal to $0$ for any input $y$:
consequently $\alpha=0$ can only be a local maximum for the MSE, with
$J_{\textrm{MSE}}(0)=\sigma_X^2$.
Moreover, by observing that the zero-mean Gaussian $x$ concentrates the useful information
around zero, it is intuitive that $J_{\textrm{MSE}}(\alpha)$ tends to decrease for values of $\alpha$ increasing from $0$,
till reaching a minimum that can be safely assumed as the optimum threshold we are looking for.
This fact is also confirmed by the shape of the MSE obtained by simulations
in Figs. \ref{fig6}-\ref{fig12}, which show that any
classical numerical solution of $J_{\textrm{BN}}^{(1,\alpha)}(\alpha) =0$
will easily converge to the optimal threshold, as well as the iterative algorithm A2
that solves the fixed-point equation in (\ref{eq71}).

More rigorously, by obvious notation equivalence, lets express
$F_{\textrm{BN}} (\alpha )$ in (\ref{eq71}) as
\begin{equation}
\label{eq72}
F_{\textrm{BN}}^{(mse)} (\alpha )=\alpha G(\alpha )=\alpha \frac{N_G (\alpha )}{D_G (\alpha
)}=\alpha \frac{\sum\limits_{m=0}^\infty {a_m e^{-\alpha ^2/2k_m }}
}{\sum\limits_{m=0}^\infty {b_m e^{-\alpha ^2/2k_m }} }.
\end{equation}
Thus, the solution of the fixed point equation in (\ref{eq71}) corresponds to $G(\alpha)=1$.
Noteworthy, as proved in the following, $G(\alpha)$ is a monotonic increasing function:
thus the solution of $G(\alpha)=1$, if it exists, is unique.
Actually, the first derivative of $G(\alpha)$ is expressed by
\begin{equation}
\label{eq74}
\begin{array}{rcl}
 G^{(1,\alpha )}(\alpha ) &=& \frac{1}{D_G^2 (\alpha )}\left[ {N_G^{(1,\alpha )}
(\alpha )D_G (\alpha )-N_G (\alpha )D_G^{(1,\alpha )} (\alpha )} \right] \\
 &=& \frac{\alpha }{D_G^2 (\alpha )}\sum\limits_{m=0}^\infty
{\sum\limits_{l=0}^\infty {\frac{b_m a_l -a_m b_l }{k_m }e^{-\frac{\alpha
^2}{2}\left( {\frac{1}{k_m }+\frac{1}{k_l }} \right)}} }, \\
 \end{array}
\end{equation}
where the terms for $m=l$ null out in the double series.
Thus, due to the symmetry when the index $m$ is interchanged with $l$,
equation (\ref{eq74}) can be rearranged as
\begin{equation}
\label{eq75}
\begin{array}{l}
G^{(1,\alpha )}(\alpha)=\frac{\alpha }{D_G^2 (\alpha
)} \sum \limits_{m=0}^\infty \sum \limits_{l=m+1}^\infty \left( \frac{b_m
a_l -a_m b_l }{k_m} \right.
\\
\qquad \qquad \qquad
\left.
+\frac{b_l a_m -a_l b_m }{k_l } \right)e^{-\frac{\alpha
^2}{2}\left( {\frac{1}{k_m }+\frac{1}{k_l }} \right)}.
\end{array}
\end{equation}
By substituting in (\ref{eq75}) the value of $a_m $, $b_m $, and $k_m $ subsumed in
(\ref{eq71}) and (\ref{eq72}), it is readily derived that
\[
\begin{array}{rcl}
G^{(1,\alpha )}(\alpha)=\frac{2\alpha }{D_G^2 (\alpha
)}\sum\limits_{m=0}^\infty {\sum\limits_{l=m+1}^\infty {\frac{\beta _m \beta
_j \sigma_X^2 }{k_m^{5/2} k_l ^{5/2}}}} ({\sigma_m^2 -\sigma_l^2})^2
\\
\cdot e^{-\frac{\alpha ^2}{2}\left( {\frac{1}{k_m }+\frac{1}{k_l }}
\right)} \ge 0
\end{array}
\]
due to the fact that all the terms inside the double series are greater than
zero. Thus, it is proved that $G(\alpha )$  [and $F_{\textrm{BN}}^{(mse)}(\alpha)$] is monotonically increasing.
Additionally, by observing that $\sigma _m^2 =m\sigma _I^2
/A+\sigma _I^2 \mathop \to \limits_{m\to \infty } \infty $, it follows that
\begin{equation}
\label{eq76}
\begin{array}{rcl}
\mathop {\lim }\limits_{\alpha \to \infty } G(\alpha )
&=&
\mathop {\lim }\limits_{\alpha \to \infty } \frac{N_G(\alpha)}{D_G(\alpha)}
= \mathop {\lim }\limits_{\alpha \to \infty } \frac{\sum\limits_{m=0}^\infty
{a_m e^{-\alpha ^2/2(\sigma _m^2 +\sigma_X^2 )}} }{\sum\limits_{m=0}^\infty
{b_m e^{-\alpha ^2/2(\sigma _m^2 +\sigma_X^2 )}} }\\
&=&\mathop {\lim }\limits_{\alpha \to \infty } \frac{a_\infty e^{-\alpha ^2/2(\sigma _\infty
^2 +\sigma_X^2 )}}{b_\infty e^{-\alpha ^2/2(\sigma _\infty ^2 +\sigma_X^2
)}}=\frac{2\sigma _\infty ^2 (\sigma _\infty ^2 +\sigma_X^2
)^{1/2}}{(\sigma _\infty ^2 +\sigma_X^2 )^{3/2}}=2
\end{array}
\end{equation}
and consequently $\mathop {\lim }\limits_{\alpha \to \infty } F_{\textrm{BN}}^{(mse)}(\alpha )=
\mathop {\lim }\limits_{\alpha \to \infty } 2\alpha =\infty$.
Thus, as shown in Fig. \ref{fig14}, the BN fixed-point problem has a different structure
with respect to the SL: $F_{\textrm{BN}}^{(mse)}(\alpha )$ is not a contraction mapping, which motivates
the use of algorithm A2 instead of algorithm A1.

It is difficult to analytically prove that $G(0)<1$, which would guarantee the existence of
the unique solution for $G(\alpha)=1$. However, it can be observed that the MSE derivative
can also be expressed as
\begin{equation}
\label{eq:derJ_BN_G}
J_{\textrm{BN}}^{(1,\alpha)}(\alpha) = \alpha^2 \left[ N_G(\alpha)-D_G(\alpha)\right].
\end{equation}
Using (\ref{eq76}), it is possible to conclude that
\begin{equation}
\label{eq:lim_derJ_BN}
\mathop {\lim }\limits_{\alpha \to \infty } J_{\textrm{BN}}^{(1,\alpha)}(\alpha) =
\mathop {\lim }\limits_{\alpha \to \infty } \alpha^2 N_G(\alpha) = 0^{+},
\end{equation}
which means that the MSE plot is an increasing function when it reaches its
asymptotic maximum $J_{\textrm{BN}}(\infty) =\sigma_{N}^2$, as intuitive
and also observable in the simulation plots.
Thus, a (unique) minimum should necessarily exist between the two
maxima $J_{\textrm{BN}}(0)=\sigma_{X}^2$ and $J_{\textrm{BN}}(\infty)=\sigma_{N}^2$.
Otherwise, the minimization problem would have no solutions, which does not make
any sense for the reasons explained before.

\bibliographystyle{IEEEtran}
\bibliography{IEEEabrv,Bibliography_abbrev}

\begin{thebibliography}{10}
\providecommand{\url}[1]{#1}
\csname url@samestyle\endcsname
\providecommand{\newblock}{\relax}
\providecommand{\bibinfo}[2]{#2}
\providecommand{\BIBentrySTDinterwordspacing}{\spaceskip=0pt\relax}
\providecommand{\BIBentryALTinterwordstretchfactor}{4}
\providecommand{\BIBentryALTinterwordspacing}{\spaceskip=\fontdimen2\font plus
\BIBentryALTinterwordstretchfactor\fontdimen3\font minus
  \fontdimen4\font\relax}
\providecommand{\BIBforeignlanguage}[2]{{%
\expandafter\ifx\csname l@#1\endcsname\relax
\typeout{** WARNING: IEEEtran.bst: No hyphenation pattern has been}%
\typeout{** loaded for the language `#1'. Using the pattern for}%
\typeout{** the default language instead.}%
\else
\language=\csname l@#1\endcsname
\fi
#2}}
\providecommand{\BIBdecl}{\relax}
\BIBdecl

\bibitem{Middleton:1}
D.~Middleton, ``Statistical-physical models of urban radio-noise environments -
  part i: Foundations,'' \emph{{IEEE} Trans. Electromagn. Compat.}, vol.
  EMC-14, no.~2, pp. 38--56 --, May 1972.

\bibitem{Middleton:1977}
------, ``Statistical-physical models of electromagnetic interference,''
  \emph{{IEEE} Trans. Electromagn. Compat.}, vol. EMC-19, no.~3, pp. 106--127,
  Aug 1977.

\bibitem{Middleton:1983}
------, ``Canonical and quasi-canonical probability models of class a
  interference,'' \emph{{IEEE} Trans. Electromagn. Compat.}, vol. EMC-25,
  no.~2, pp. 76--106, May 1983.

\bibitem{Berry:1981}
L.~Berry, ``Understanding middleton's canonical formula for class a noise,''
  \emph{{IEEE} Trans. Electromagn. Compat.}, vol. EMC-23, no.~4, pp. 337--344,
  Nov 1981.

\bibitem{Middleton:1999}
D.~Middleton, ``Non-gaussian noise models in signal processing for
  telecommunications: New methods and results for class a and class b noise
  models,'' \emph{{IEEE} Trans. Inf. Theory}, vol.~45, no.~4, pp. 1129--1149,
  May 1999.

\bibitem{Middleton:1979}
------, ``Procedures for determining the parameters of the first-order
  canonical models of class a and class b electromagnetic interference [10],''
  \emph{{IEEE} Trans. Electromagn. Compat.}, vol. EMC-21, no.~3, pp. 190--208,
  Aug 1979.

\bibitem{Zabin:1989}
S.~Zabin and H.~Poor, ``Parameter estimation for middleton class a interference
  processes,'' \emph{{IEEE} Trans. Commun.}, vol.~37, no.~10, pp. 1042--1051,
  1989.

\bibitem{Zabin:1991}
------, ``Efficient estimation of class a noise parameters via the em
  algorithm,'' \emph{{IEEE} Trans. Inf. Theory}, vol.~37, no.~1, pp. 60--72,
  1991.

\bibitem{Blackard:1993}
K.~Blackard, T.~Rappaport, and C.~Bostian, ``Measurements and models of radio
  frequency impulsive noise for indoor wireless communications,'' \emph{{IEEE}
  J. Sel. Areas Commun.}, vol.~11, no.~7, pp. 991--1001, 1993.

\bibitem{Zhong:2007}
Y.-Z. Jiang, X.~lin Hu, X.~Kai, and Z.~Qi, ``Bayesian estimation of class a
  noise parameters with hidden channel states,'' in \emph{Proc. {IEEE} Int.
  Symp. on Power Line Comm. and its Appl.({ISPLC}'07)}, march 2007, pp. 2 --4.

\bibitem{Rappaport:1966}
S.~Rappaport and L.~Kurz, ``An optimal nonlinear detector for digital data
  transmission through non-gaussian channels,'' \emph{{IEEE} Trans. Commun.},
  vol.~14, no.~3, pp. 266--274, Jun 1966.

\bibitem{Spaulding:1977}
A.~Spaulding and D.~Middleton, ``Optimum reception in an impulsive interference
  environment--part i: Coherent detection,'' \emph{{IEEE} Trans. Commun.},
  vol.~25, no.~9, pp. 910--923, Sep 1977.

\bibitem{Spaulding:1985}
A.~Spaulding, ``Locally optimum and suboptimum detector performance in a
  non-gaussian interference environment,'' \emph{{IEEE} Trans. Commun.},
  vol.~33, no.~6, pp. 509--517, Jun 1985.

\bibitem{Middleton:1995}
D.~Middleton, ``Threshold detection in correlated non-gaussian noise fields,''
  \emph{{IEEE} Trans. Inf. Theory}, vol.~41, no.~4, pp. 976--1000, Jul 1995.

\bibitem{Stein:1995}
D.~Stein, ``Detection of random signals in gaussian mixture noise,''
  \emph{{IEEE} Trans. Inf. Theory}, vol.~41, no.~6, pp. 1788--1801, 1995.

\bibitem{Maras:2003}
A.~Maras, ``Adaptive nonparametric locally optimum bayes detection in additive
  non-gaussian noise,'' \emph{{IEEE} Trans. Inf. Theory}, vol.~49, no.~1, pp.
  204--220, Jan 2003.

\bibitem{Zhidkov:2006}
S.~Zhidkov, ``Performance analysis and optimization of ofdm receiver with
  blanking nonlinearity in impulsive noise environment,'' \emph{{IEEE} Trans.
  Veh. Technol.}, vol.~55, no.~1, pp. 234--242, Jan 2006.

\bibitem{Mayers:2003}
E.~S{\"u}li and D.~Mayers, \emph{An introduction to numerical analysis}.\hskip
  1em plus 0.5em minus 0.4em\relax Cambridge Univ Pr, 2003.

\bibitem{Papoulis:1991}
A.~Papoulis, \emph{Probability, Random Variables, and Stochastic
  Processes}.\hskip 1em plus 0.5em minus 0.4em\relax McGraw-Hill, 1991.

\bibitem{Kyees:1995}
P.~Kyees, R.~McConnell, and K.~Sistanizadeh, ``Adsl: a new twisted-pair access
  to the information highway,'' \emph{{IEEE} Commun. Mag.}, vol.~33, no.~4, pp.
  52--60, Apr 1995.

\bibitem{Pavlidou:2003}
N.~Pavlidou, A.~Han~Vinck, J.~Yazdani, and B.~Honary, ``Power line
  communications: state of the art and future trends,'' \emph{Communications
  Magazine, IEEE}, vol.~41, no.~4, pp. 34--40, 2003.

\bibitem{Henkel:1995}
W.~Henkel, T.~Kessler, and H.~Chung, ``Coded 64-cap adsl in an impulse-noise
  environment-modeling of impulse noise and first simulation results,''
  \emph{{IEEE} J. Sel. Areas Commun.}, vol.~13, no.~9, pp. 1611--1621, 1995.

\bibitem{Zimmermann:2002}
M.~Zimmermann and K.~Dostert, ``Analysis and modeling of impulsive noise in
  broad-band powerline communications,'' \emph{Electromagnetic Compatibility,
  IEEE Transactions on}, vol.~44, no.~1, pp. 249--258, 2002.

\bibitem{Ma:2005}
Y.~Ma, P.~So, and E.~Gunawan, ``Performance analysis of ofdm systems for
  broadband power line communications under impulsive noise and multipath
  effects,'' \emph{Power Delivery, IEEE Transactions on}, vol.~20, no.~2, pp.
  674--682, 2005.

\bibitem{Nassar:2011}
M.~Nassar, K.~Gulati, Y.~Mortazavi, and B.~Evans, ``Statistical modeling of
  asynchronous impulsive noise in powerline communication networks,'' in
  \emph{Proc. {IEEE} Int. Global Commun. Conf.}, December 2011, pp. 1--6.

\bibitem{Zhidkov:2008}
S.~Zhidkov, ``Analysis and comparison of several simple impulsive noise
  mitigation schemes for ofdm receivers,'' \emph{{IEEE} Trans. Commun.},
  vol.~56, no.~1, pp. 5--9, Jan 2008.

\bibitem{Guo:2005}
D.~Guo, S.~Shamai~(Shitz), and S.~Verdù, ``Mutual information and minimum
  mean-square error in gaussian channels,'' \emph{{IEEE} Trans. Inf. Theory},
  vol.~51, no.~4, pp. 1261--1283, Apr 2005.

\bibitem{Kay:1993}
S.~M. Kay, \emph{Fundamentals of Statistical Signal Processing. Vol. 1,
  Estimation Theory}.\hskip 1em plus 0.5em minus 0.4em\relax Prentice-Hall,
  1993.

\bibitem{Proakis:1987}
J.~Proakis, \emph{Digital communications}.\hskip 1em plus 0.5em minus
  0.4em\relax McGraw-hill, 1987.

\bibitem{Banelli:2011IT}
\BIBentryALTinterwordspacing
P.~Banelli, ``Another useful theorem for non-linear transformations of gaussian
  random variables,'' \emph{arXiv:1111.5950v1 (cs.IT)}, no. submitted to IEEE
  Trans. on Inf. Theory, pp. 1--7, November 2011. [Online]. Available:
  \url{http://arxiv.org/abs/1111.5950}
\BIBentrySTDinterwordspacing

\bibitem{Rowe:1982}
H.~E. Rowe, ``Memoryless nonlinearities with gaussian inputs: Elementary
  results,'' \emph{Bell Syst. Tech. J.}, vol.~61, no.~7, pp. 1519--1525, Sep
  1982.

\bibitem{Bussgang:1952}
\BIBentryALTinterwordspacing
J.~J. Bussgang, ``Crosscorrelation functions of amplitude-distorted gaussian
  signals,'' \emph{M.I.T. RLE Technical Report}, no. 216, pp. 1 --14, march
  1952. [Online]. Available: \url{http://hdl.handle.net/1721.1/4847}
\BIBentrySTDinterwordspacing

\bibitem{Baum:1957}
R.~Baum, ``The correlation function of smoothly limited guassian noise,''
  \emph{IRE Trans. Inf. Theory}, vol. IT-3, pp. 193--197, Sep 1957.

\bibitem{Davenport:1958}
W.~B. Davenport~Jr. and W.~L. Root, \emph{An Introduction to the Theory of
  Random Signals and Noise}.\hskip 1em plus 0.5em minus 0.4em\relax Mc Graw
  Hill, 1958.

\bibitem{Banelli:2000}
P.~Banelli and S.~Cacopardi, ``Theoretical analysis and performance of ofdm
  signals in nonlinear awgn channels,'' \emph{{IEEE} Trans. Commun.}, vol.~48,
  no.~3, pp. 430--441, Mar 2000.

\bibitem{Interference:1}
K.~Gulati, M.~Nassar, A.~Chopra, N.~Ben~Okafor, M.~DeYoung, N.~Aghasadeghi,
  A.~Sujeeth, and B.~L. Evans, \emph{Interference Modeling and Mitigation
  Toolbox 1.6, for Matlab}, ESP Laboratory, ECE Dept., Univ. of Texas at
  Austin, Oct 2011.

\end{thebibliography}

\begin{biography}{Paolo Banelli}
Paolo Banelli received the Laurea degree in electronics engineering and the Ph.D. degree in telecommunications
from the University of Perugia, Italy, in 1993 and 1998, respectively.
In 2005, he was appointed Associate Professor at the Department of Electronic and Information Engineering (DIEI),
University of Perugia, where he has been an Assistant Professor since 1998.
In 2001, he joined as a visiting researcher the SpinComm group,
lead by Prof. G.B. Giannakis, at the Electrical and Computer Engineering Department,
University of Minnesota, Minneapolis.
His research interests mainly focused on signal processing for wireless communications,
with emphasis on multicarrier transmissions, and more recently on signal processing
for biomedical applications, with emphasis on electrocardiography and medical ultrasounds.
He has been serving as a reviewer for several technical journals,
and as technical program committee member of leading international conferences
on signal processing and telecommunications.
In 2009, he was a General Co-Chair of the IEEE International Symposium
on Signal Processing Advances for Wireless Communications.
In December 2010 he has been elected as a member of the SPCOM Technical Committee
of the IEEE Signal Processing Society, where he serves since January 2011.
He is a co-founder (2010) and scientific co-director of ICT4Life
(http://www.ict4life.it), a spin-off company of University of Perugia.
\end{biography}
\vfill
\vfill
\vfill






\end{document}